# Critical misalignments in climate pledges reveal imbalanced sustainable development pathways

Larosa, F.[1,*], Mallor, F.[1], Hoyas, S.[2], Conejero, J.A.[2], Garcia–Martinez, J.[3], Fuso Nerini, F.[4], Vinuesa, R.[1],

[1] FLOW, Engineering Mechanics, KTH Royal Institute of Technology, Stockholm, Sweden
2 Instituto Universitario de Matemática Pura y Aplicada, Universitat Politècnica de València, Valencia, Spain
3 Departamento de Química Inorgánica, Universidad de Alicante, Alicante, Spain
4 KTH Climate Action Center and Division of Energy Systems, KTH Royal Institute of Technology, Stockholm, Sweden

* Corresponding author: larosa@kth.se

## ABSTRACT

We explore the integration of climate action and Sustainable Development Goals (SDGs) in nationally determined contributions (NDCs), revealing persistent synergies and trade-offs across income groups. While high-income countries emphasize systemic challenges like health (SDG3) and inequality (SDG10), low-income nations prioritize the water-energy-food nexus (SDGs 6-7-12) and natural resource management (SDG15) due to vulnerabilities to climate impacts. Harnessing an innovative artificial intelligence routine, we discuss what these diverging development trajectories imply for the Paris Agreement and the 2030 Agenda for sustainable development in terms of global inequality, the climate and sustainable finance flows and multilateral governance.

## MAIN

The 29th Conference of Parties (COP29) in Baku earlier this year closed with conflicting feelings about the future of climate action. Countries conveyed in Azerbaijan and reached unprecedented agreements on climate finance and carbon markets, but observers agree that huge gaps remain between commitments and needs. Amid converging crises, the planet is dangerously set to warm by 3.1°C if "current policies" are implemented[1]. To move from pledges to implementation, COP29 in Baku (Azerbaijan) and COP30 in Belem (Brazil) develop in continuity with its predecessor under the COP Presidencies Troika. COP28 in Dubai in December 2023 closed the first-ever Global Stocktake (GST) process: a global and comprehensive evaluation of collective progress towards meeting the 2015 Paris Agreement. The GST did not simply identify and address critical barriers and existing gaps in the climate governance landscape. It also called for a stronger alignment between climate action and sustainable development objectives[2]. While there is convergence towards the idea that the climate and sustainability concepts are inseparable[3], stakeholders still differ in their understandings of how the 17 Sustainable Development Goals (SDGs) can be operationalised under the strict temperature 1.5°C target set at COP21. The challenge ahead is complex: a mere 12 percent of SDG targets are on track to be achieved by 2030[4] and the planet is showing concerning signs of distress as six out of nine planetary boundaries were proven to be transgressed[5]. To move along just transition pathways, fossil fuels need to be ordinately phased-out, while policy-makers must ensure that low-carbon technologies scale-up quickly, benefitting also the most vulnerable groups[6]. Policy must foresee, assess and avoid unintended environmental consequences of new production and consumption models[7], ruling over the new governance of a sustainable society[8].

As multiple and simultaneous policy interventions are needed to accomplish sustainability goals, a fertile research stream has explored the interlinkages between the 17 SDGs[9] to support policy-makers in fulfilling the 2030 Agenda. While synergies typically outweigh trade-offs, efforts to meet specific goals - including SDG13, climate action - limit or reduce progress in other domains[10]. Evidence of SDG interlinkages and their nature is gathered through automatic[11] or manual[10] large-scale reviews of academic literature and reports[9]. Bird-eye views of the SDG interaction landscape[9], context-relevant analyses[12] and area-specific studies[13,14] enriched a holistic understanding of systemic sustainability barriers. While useful, published evidence focuses on past or current interlinkages falling short in supporting decision-makers with actionable knowledge. Some studies also suffer from low replicability as they require assumptions subject to certain degrees of subjectivity[15]. Due to the time-consuming procedures behind this work, evidence production may be too slow to support action. In the climate-action domain, where trade-offs have proved to be serious threats to other SDGs[13], the analysis of win-win solutions to advance adaptation, mitigation and sustainable development must be timely, bespoke and accessible to policy-makers. Past research shows that siloed plans for climate change adaptation and mitigation can undermine efforts to achieve other SDGs, unless the climate agenda is situated within the overall SDG framework. However, only a handful of the national plans submitted by Parties under the Paris Agreement explicitly take into account broader sustainable development outcomes[13]. We focus on the interaction between climate and other SDGs using the Nationally Determined Contributions (NDCs) as key programmatic documents of parties' future adaptation and mitigation plans. We do so to uncover opportunities for stronger alignment between climate and sustainability agendas and to critically examine disconnected areas in light of the "common but differentiated responsibilities and respective capabilities" principles (CBDR-RC) in the UN Framework Convention on Climate Change (UNFCCC). The identification of misalignments between the Paris Agreement and the needs to advance the 2030 Agenda for Sustainable Development is instrumental to policy coordination as advocated by the UN Expert Group on Climate and SDG Synergy[4], the European Commission[16] and the 6th Assessment Report of the Intergovernmental Panel on Climate Change (IPCC)[17].

The need for timely and continuous assessment of consistencies across the two intertwined agendas has been flagged by international agencies[18], organisations[19] and research institutions[20]. Academic studies have also tackled the type and breadth of interactions between climate and the SDGs[21] using descriptive[13], model-based[22] and data-driven[23] approaches. While indispensable in identifying past and present relationships across different sustainability dimensions, these works often rely on stringent assumptions and limited views regarding some SDGs (especially the social ones)[24]. We track alignment between the NDCs and SDGs contributing to the policy landscape in two ways. First, we go beyond the simple identification of climate-SDG links and we characterize them based on their impact on climate adaptation and mitigation. Second, we reflect upon what current and future research can do to tackle critical misalignment in order to move along climate-compatible sustainable development pathways. As parties are expected to submit new, more ambitious and stronger NDCs by the first months of 2025, the learning of past submissions constitutes an important building block to advance global sustainable development, negotiations over climate finance allocation and needs-based technological transfers.

NDCs are programmatic documents: they uncover what countries *aim and plan to do*, rather than what countries have been actually doing to reach the Paris target without compromising the sustainability of their economy. Submitted to the UNFCCC, the NDCs outline national goals, priorities and intentions to unfold climate action efforts. As such, they are political documents with clearly defined goals. While mostly focused on the climate mitigation side, Parties started including an adaptation component since the 2020-2021 submission cycle to strengthen their ambition and to widen the spectrum of potential synergies with the SDGs. The discourse analysis of the NDCs and the focus on them as narratives has been helpful to assess the credibility of pledges[25] and the countries' differentiated responsibilities in addressing the climate issue[26]. Given that NDCs serve as a primary instrument for long-term international cooperation and negotiation, ensuring their comparability is essential. However, the diversity of these contributions presents a significant challenge for qualitative discourse analysis, complicating consistent comparisons across contexts[27]. Most of the efforts look at launching tools to guide countries in building their future pledges. However, a bird-eye view on the NDCs informs global governance structures and opens a reflection on the current procedural mechanisms to pledge for climate action. To support human expertise, we process the textual corpus with artificial intelligence (AI) methods, which act as a resource for first-pass assessment on how the NDCs could be harmonised with the SDGs, responding to the call to action to put the GST in place. We include 158 Parties (out of 195 parties by UNFCCC Registry), representing per-capita high- (50 parties), medium- (36) and low-emitting nations (72) in per capita terms. Updated (up to the second submission) NDC versions are included.

**A WEAK ALIGNMENT**

The majority of countries considered (55.1%, 87 parties) do not explicitly mention or refer to the SDGs in their NDCs (Figure 1a). While this misalignment is concerning for every country as the SDGs form a universal agenda, those countries that experience higher vulnerability and lower readiness to climate change face higher hurdles. As sustainability issues and the SDGs may be implicitly discussed or concealed in the nuances of the text, relying solely on explicit terms or keywords may not capture their full presence. We implement an AI-based procedure using the Google Research Large Language Model (LLM) Gemini 1.0[28] to classify the paragraphs extracted from the NDCs into one or more SDGs, based on the implicit meaning conveyed by the text (Figure 1b). The capability of AI to extract non-immediate insights from texts supports the elicitation of critical information about the relationship between vulnerability to climate change and the importance that countries assign to the SDGs by accounting for them explicitly in their framing.

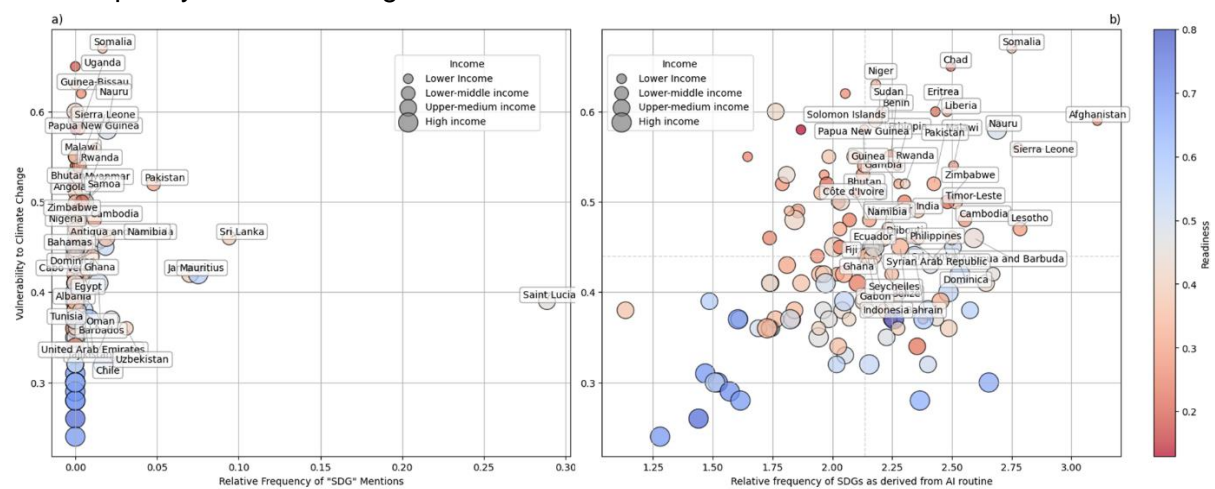

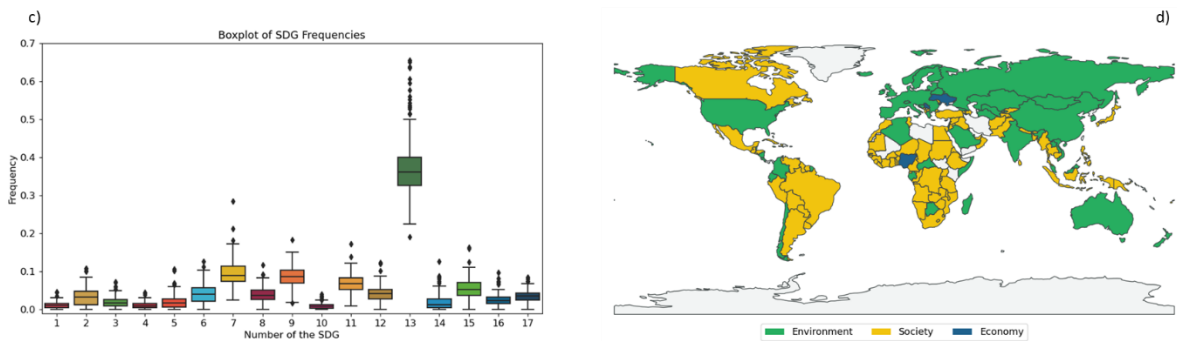

Figure 1. **a)** The search for explicitly mentioned SDGs reveal a significant alignment gap with very little text (1.01%) expanding on the two-way relationship between climate actions and sustainable development. **b)** The AI-powered routine classifies the NDCs into the appropriate SDGs based on the meaning conveyed in paragraph-level texts. Over one fourth of the sample (41 countries) is located in the upper right-hand quadrant. **c)** The weak alignment between climate policy and the 2030 Agenda for Sustainable Development is mostly related to social aspects (education, SDG4 above all). **d)** SDGs are grouped per area (environment, society and economy) and depict a polarized world with environmental factors mostly mentioned in higher and upper-middle income countries.

Specifically, the positive correlation between the SDG-relevant text and vulnerability to climate change suggests that some sense of urgency is perceived. Using the income classes by the World Bank, we observe this relationship especially in lower-middle (39%) or lower-income (34%) countries. Among the high-income countries making stronger reference to the SDGs in their NDCs, fossil fuel dependent economies (e.g., Saudi Arabia, Chile) use the SDGs as a framework to plan a just transition as they need to restructure their energy systems. Once grouped as in Norström et al.[29], the SDGs depict a polarized world where SDG13 (climate action) is widely tackled by the Global North and social issues such as poverty and hunger are prioritized in the Global South. In agreement with previous literature[13], this representation indicates that Parties submitting the NDCs tailor their policies differently and still poorly embrace a holistic view of the sustainable development agenda. This diverging position of countries concerning SDG13 also suggests that emerging and developing economies – more than high-income ones – recognize climate adaptation and mitigation as non-exclusive, but integrated into a broader sustainable development domestic strategy (Figure 1d).

This has huge implications for the architecture of climate and sustainable finance agreements. The identification of key SDG-NDCs links can redefine both quality and quantity of financial flows needed to support vulnerable countries over the transition. As different intertwined challenges disproportionally affect low and middle income countries (LMIC) debt service often takes precedence over domestic investments to build resilience[30]. Different financial instruments (grants, concessional loans, bonds, insurance, funds and swaps) to promote climate-compatible development agendas have generated debt distress for 79 countries, sixty of which are highly vulnerable to climate change[31]. This is partly because Debt Sustainability Analyses have consistently overlooked climate change. Even when climate change and development are integrated – for example in the World Bank's Group Country Climate and Development Reports – the pathways identified do not consider synergic opportunities, but treat every sector as independent. However, the study of interlinkages paves the road to a more holistic outlook and can inform if and how different financial instruments can be used to mitigate or eliminate pressing challenges. Persistent co-occurrences among SDGs (Figure 2a) are a key example of these unexplored opportunities. Forest management, afforestation and biodiversity protection (SDG15) are priorities especially for Sub-Saharan African Parties, but only when linked to agricultural development and land use changes. Debt-for-nature swaps may be valid options to simultaneously address raising interest rates on debt, biodiversity loss and climate harms.

Across all the SDGs, the *energy-infrastructure-community* (*SDG7-SDG9-SDG11*) nexus is the most persistent (Figure 2a): it builds on the debate around the just energy transition, but it also expands on the structural requirements that prevent risks for cities and urban settlements. While in LMICs the focus is on needs and climate-induced impacts, upper-middle and high-income countries tackle the implementation side anchoring their narratives on means to meet the pledges. This difference raises questions about the role of the NDCs in global governance. As parties leverage on different angles to describe their pledges, the definition of a unified and shared scope and format for the NDCs may be instrumental to orderly achieving the Paris Agreement target. Furthermore, it can boost comparability creating a common benchmark of intentions against which progress can be tangibly measured. For example, the NDCs acknowledge that inequality considerations exist (Figure 2b). This is a pivotal issue as policies may create imbalances especially towards the most vulnerable groups and lead to domestic conflicts and tensions (SDG16). Inequality may also be a product of other policies: sustainable growth (SDG8), gender (SDG5) and sustainable communities (SDG11). Hence, the narrative built around SDG10 and the way Parties develop their narratives, profoundly change the message. Despite being loosely mentioned, inequality is a pervasive topic that strongly links to all SDGs (Figure 2b) and a qualitative, critical and discourse analysis becomes essential.

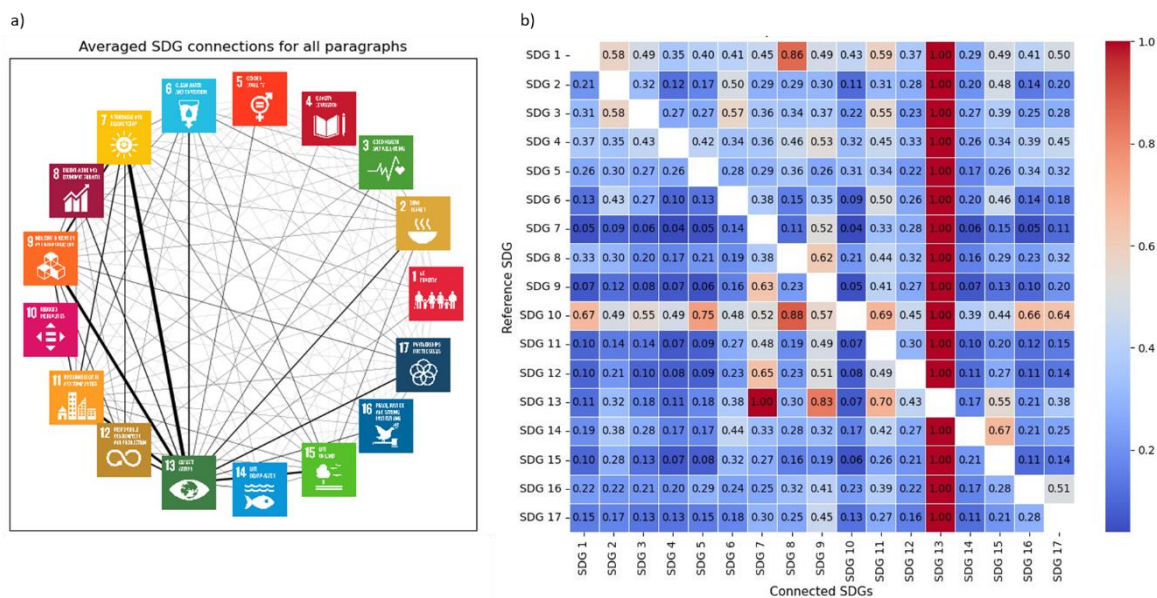

Figure 2. **a)** Connections are averaged per number of paragraphs within each NDC to ensure comparability and over the total number of countries. The *energy-infrastructure-communities* and the *infrastructures-partnerships* nexuses are the most stable as represented by the thickness of connections. **b)** Co-occurrences heat map displays the relative weight across all SDGs. Each score describes how relevant each connection is (on the X axis) with respect to each specific SDG (from the Y axis).

## USING NDC-SDG INTERLINKAGES AS A SYSTEM TRANSITION FRAMEWORK

The alignment between NDCs and SDGs also depends on the framing as parties unevenly present SDG interlinkages both in frequency and tone. Based on the Party's characteristics and its related risks, links between climate policy and other sustainable development dimensions may have positive, neutral or negative effects on domestic adaptation and mitigation strategies, hence leading to explicit synergies and trade-offs. The NDC-SDG interlinkages as a methodological approach can be transformed into a robust policy tool to assess system transition dynamics. The critical analysis of positive and negative spillovers describes a complex system of interactions with isolated, connected and emerging clusters of issues and stakeholders. These assessments pose system transitions at the heart of transformative pathways and the text of the NDCs reveal ways to unlock policies which support a climate-resilient sustainable development. Only half of the text in the NDC is non-neutral (51.09%) and typically positive (synergic) outweighs negative (trade-off) links between SDG13 and other SDGs. Trade-offs require more elaboration (Figure 3a), in-depth knowledge and context-relevant assessments. Countries lacking adequate capacity may be disadvantaged in addressing these issues. The support from non-governmental bodies to equip countries with updated knowledge raises the ambition to assess both synergies and trade-offs. However, a more structured focus on finding negative interlinkages may prevent unintended consequences. In the future, the NDCs may be used to elicit from parties which development actions contribute or detriment to climate adaptation and mitigation. This would help building in-country capacity and knowledge for the future and would enhance comparability across different documents. It would also enhance the credibility of pledges. By treating the NDC-SDG ecosystem as a complex network of interactions, several control mechanisms can be designed to evaluate progress. An example of a system transition dynamic is the aforementioned *energy-infrastructure-community* nexus. LMICs are more explicit in both synergies and trade-offs than upper-middle and high-income countries (Figure 3b) as they reinforce the idea that planning climate actions may be constrained by financial resources availability. On the trade-offs side, LMICs highlight that within the *SDG7-SDG9-SDG11* nexus (Figure 3c), a simultaneous expansion of infrastructure networks, energy grids and low-carbon technologies is a pre-condition for the system transition to happen. This explicit call for more holistic development planning is consistent with the idea that there is no "one-size-fits-all" solution as feasibility and effectiveness are highly heterogeneous. In fact, high-income countries tackle the positive links between SDG7, SDG9 and SDG11 more frequently than LMICs, but they anchor their narrative on benefits for climate adaptation and mitigation: they target energy efficiency in key sectors (e.g., industry, building and land transportation – Saudi Arabia) and mitigation reduction policy via electrification and fuel substitution (e.g., from oil to hydrogen in transport - Uruguay).

The framework welcomes and explores whether the identified trade-offs encompass both adaptation and mitigation. Issues related to natural resources conservation and management (SDG6, SDG15 and SDG2) and impacts on livelihoods are particularly relevant when it comes to climate adaptation in LMICs, but they constitute key assets to reduce greenhouse gas emissions (mitigation) in higher-income Parties. This new web of climate actions is shaped by a heterogeneity in skills, interests and capabilities and LMICs lead the way in embracing complexity across adaptation and mitigation, social, environmental and economic dimensions. Sub-Saharan African parties for example (e.g., Sierra Leone and Togo) are among the most vocal in flagging the need to protect communities depending on freshwater bodies and fishery from climate impacts; South Asian parties (e.g., Tajikistan) introduce the need to expand adaptation measures for rural communities through climate finance. These links are made explicit in NDCs submitted primarily after the COVID-19 pandemic when sustainable development progress suffered from widespread halts.

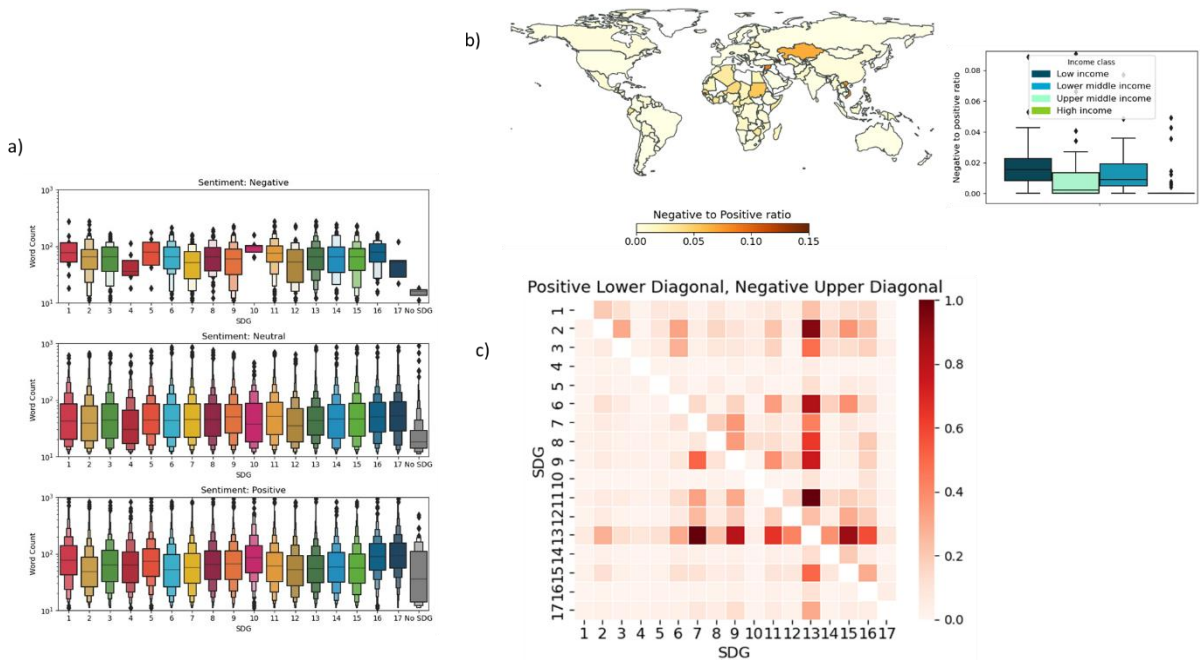

**Figure 3. a)** The largest variability in SDG distribution per word count and tone is visible in the negative paragraphs with socio-economic SDGs as discoverable in longer chunks. SDG4 confirms itself as poorly represented in the NDCs; **b)** Parties are assigned colored per intensity of negative/positive tone ratio: the darker the color, the more negatively each party frames its NDC with respect to climate adaptation and mitigation; **c)** Heatmap illustrating the normalized average frequency of SDG-to-SDG co-occurrences in positive (lower triangle) and negative (upper triangle) paragraphs. For each SDG pair, we calculate the frequency of co-occurrence per paragraph in each country, average these frequencies across all countries, and normalize the values within each sentiment class.

The study of the NDCs-SDGs interactions as a system transition framework can unlock 1USD trillion investment bringing more projects online faster. It is the call to action issued by the Industrial Transition Accelerator (ITA) at COP29, a collaboration of global business leaders and investors. By leveraging on synergic opportunities in key sectors of the economy, a new green industrial policy can arise simultaneously stimulating sustainable growth and climate action.

**BUILD PEACEFUL AND STRONG INSTITUTIONS THROUGH CLIMATE ACTION**

Synergic interdependencies between the SDGs and climate adaptation and mitigation are offset by the existence of trade-offs between different agendas. Strongly recognized and highlighted by LMICs (Figure 4), these negative interactions are mitigated when cooperation, collaboration and trusted multilateral institutions are in place. Multilateralism is at the heart of the United Nations system: adopted in 1992, the UNFCCC conveys the Conference of Parties (COPs) to negotiate over and operationalize actions to address climate change. Multilateral institutions are typically chosen not just to host, but also to manage complex issues. The first day of COP29 in Baku in November 2024 approved an UN-administered global carbon market for example, marking a breakthrough decision. Opening COP29 in Baku earlier this year, UN climate change executive secretary Simon Stiell said that the UNFCCC process remains "the only place we have to address the rampant climate crisis"[32]. At the same time, the failures of multilateralism in delivering the promise of a prosperous and sustainable future cannot be left unaddressed. The NDCs reveal multilateralism's weaknesses when reporting the current and future trade-offs between the Paris Agreement and the 2030 Agenda for Sustainable Development. Multilateral institutions can facilitate and propose scalable solutions through flagship and multi-year programs and the NDCs provide evidence of the political priorities they should pursue for the benefit of all.

First, a peaceful geopolitical landscape favors cooperation and collaboration to reduce and remove roadblocks to a climate-smart sustainable development. Conflict-affected areas (e.g., Syria and Sudan) are also translating their challenges into the NDCs, revealing how relevant peaceful institutions (SDG16) are to foster climate action. Sabotages and attacks for example to critical infrastructure (including dams, irrigation networks and oil and gas fields) prevent the full alignment of domestic priorities and climate change adaptation and mitigation objectives (e.g., Syria's NDC explicitly mentions how "random extraction of crude oil [...] caused a significant environmental contamination"). The current multilateral governance rooted in the UN General Assembly with a five permanent Security Council mobilized to direct and regulate over global events is showing its pitfalls. The urgency to adapt the model designed after the Second World War to the need for a hyperconnected and challenged world has never been stronger. Climate action can act as guiding framework for such a reform: as countries share similar needs and experience comparable trade-offs, new alliances can be formed.

Second, countries report in their NDCs that tensions and trade-offs exist also within their borders between different agroecological areas: native forest coverage – for example - affects land use changes and discourage agricultural practices (i.e., the Democratic Republic of the Congo) which may also alter water cycles. Whenever coastal areas exist, climate changes on agricultural land have led to overfishing with consequences on food security. Challenges in supporting sustainable urban communities (SDG11), climate-smart agriculture (SDG2) for all and the adoption of low-carbon technologies (SDG7) are of primary concern, especially in low infrastructure countries' NDCs. These unstable equilibria can be corrected and supported by strong multilateral institutions which value certain areas as universally valuable, calling the entire international community to preserve them. The Cali Fund launched during the last biodiversity summit, COP16, exemplifies this proposal: the economic resources collected from the use of genetic codes of organisms shared digitally, will be allocated to Indigenous People either directly or through governments, recognizing their role in protecting biodiversity on the planet[33].

Third, the NDCs further reveal other trade-off mitigation channels that multilateral institutions shall explore to promote alignment with other SDGs. Parties detail climate risks affecting households, including sea level rise, floods and droughts. Extreme events increase the vulnerability of already vulnerable groups (e.g. Cameroon), especially in those countries where rapid urbanization processes have created chaotic developments and inadequate infrastructures (e.g., Sierra Leone). Agriculture-dependent countries, mostly in Sub-Saharan Africa (Figure 4a), witness unpredictable and quick changes that affect yield production and income generation opportunities for rural areas. Insurance schemes are envisaged as protection mechanisms for smallholder farmers (e.g., Malawi), but very few are actually in place and their development depends on climate finance programs, which are in turn widely supported by multilateral schemes. Multilateral institutions are the best equipped to support climate knowledge-sharing tools (i.e., climate services) and to design capacity-building programmes in national meteorological services. The Global Framework for Climate Services, for example, assists decision-makers in assessing and protecting against climate-related risks and often deploys a sector-based approach that strengthens the local economies. The integration of climate services in national weather and meteorological services can be encouraged through capacity-building activities that multilateral organisations can promote. Through its climate services toolkit, the World Meteorological Organisation is pioneering the provision of high quality and bespoke products and services, but challenges remain[34]. The analysis of trade-offs between climate and sustainable development can act as a useful socio-economic and environmental benefit assessment tool. Those dimensions with wider mismatch can be expanded, targeting efforts and investments to improve preparedness at the local level. To achieve this goal, regional cooperation is essential[34] and multilateral institutions act as liaison nodes in complex ecosystems of interests[35].

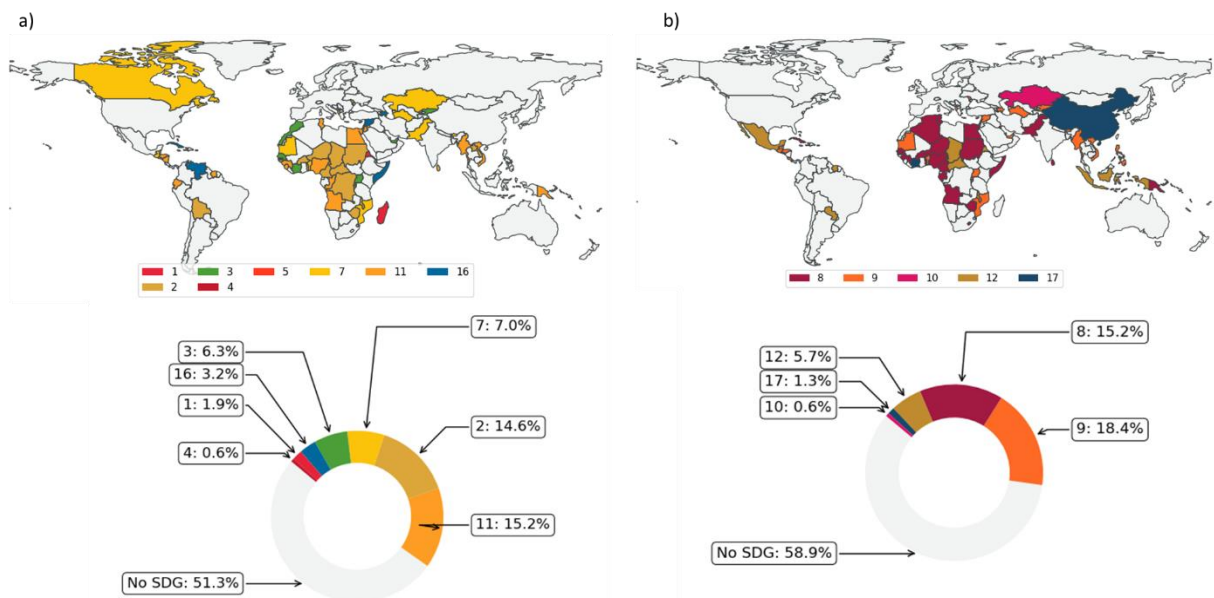

**Figure 4. An in-depth look at trade-offs across society and economy SDGs. a)** Societal SDGs are widespread across different income classes mostly due to the energy transition. As climate changes, countries flag the need to train new professionals and qualified personnel to support climate adaptation and mitigation strategies (e.g., Burundi). **b)** Economic SDGs reveal that macroeconomic issues are at the heart of climate action. International cooperation is called out by China, which highlights the need to counteract polarization and decentralized governance structure.

Finally, the NDCs bring forward climate justice by detailing macroeconomic synergies and trade-offs of selected measures. The urgent need for a rapid and scalable energy transition (SDG 7) is a shared priority for both low and high-income countries. However, challenges are different as some countries face roadblocks in financing their infrastructure due to high cost of capital ("lack of access to cheap loans for low-carbon projects" – Kazakhstan); in some cases, the transition is perceived as a gradual diversification, but with no sudden halt of exploration activities to fight against price raises ("The Ontario government is also expanding access to natural gas across the province to help keep the cost of energy low for families, businesses and farmers" – Canada). Balance in the energy mix is also an important risk factor (e.g., Moldova) as natural disasters – such as spontaneous fires – may stress power supply without solid backup options. Given that LMICs frequently frame trade-offs between SDG7 and climate adaptation and mitigation in terms of climate impacts and shortage of technical options, climate justice must consider a consistent and stable transfer mechanism from wealthy and prepared Parties to those challenged by these hurdles. The findings underscore the critical importance of modeling both physical and transition risks associated with climate change. Scenario-based transition pathways can identify regressive effects on specific sectors within the country. These analyses can wave those concerns related to the impacts of measures such as carbon pricing ("Social impacts, such as unemployment generated by self-charging electric vehicle stations" – Bahamas; *"The introduction of carbon pricing can lead to an increase in prices for fuel and energy resources and dependent services, and to a significant increase in inflation, which in turn can significantly worsen the welfare of the population"* - Kazakhstan). These narratives are particularly strong for high per-capita emitting countries and for fossil-dependent economies.

The COP29 climate negotiations concerning a new target for climate finance and a global carbon pricing mechanism take into account issues of redistribution and equity. However, as the new NDCs will need bolder, more ambitious and concrete targets to drive deep emission cuts and promote sector-specific adaptation measures, countries must assess the implications of their pledges ex ante, preserving social cohesion to avoid polarized views on climate action. Our analysis emphasizes how relevant social protection measures and targeted, time-limited economic supports will be especially in LMICs and economies with fossil-dominated consumption and production modes. Furthermore, the analysis of the first and updated NDCs reveals that – when designing such complex policy processes – inconsistencies may arise at the domestic level. Supranational and multilateral institutions can help ensure policy coherence as countries are often confronted with domestic legislation requiring updates (*"Policy inconsistencies: Actual Electricity Act of 1956 does not allow independent power producers (IPPs) to sell to the national grid. This is a major barrier to the use of renewable sources"* – Bahamas) or perceived unfair mechanisms (*"The Unilateral Coercive Measures (MCUs) ignore not only legal norms, but are also a blatant violation of ethical principles and can be viewed as a crime against humanity"* – Venezuela; *"In this context, the country has achieved important milestones, despite the prevalence of extreme challenges inherited from the condition of being a Small Island Developing State (SIDS) under a sturdy economic, commercial and financial blockade imposed by the United States of America, which has been intensified to record levels in the last few years by the Trump Administration – the latter being the main obstacle to the achievement of major progress when facing climate change and national development"* - Cuba).

**TOWARDS A NEEDS-BASED AGENDA FOR THE FUTURE**
The call for multilateral institutions to promote and monitor the integration of climate and sustainable development agendas aligns with the need to identify and represent local priorities. While the responsibilities of climate change are univocally imputable to the release of greenhouse gas emissions to which many parties contributed, impacts are very heterogeneous and unevenly distributed. Therefore, countries revise their pledges according to their national priorities. Negotiations, then, make similarities, shared interests and common goals explicit. We detect persistent synergies and trade-offs that survive across time and space and we observe that LMICs and higher-income countries converge with important exceptions. On the synergy side (Figure 5a), our findings align with previous literature[13] and highlight marked opportunities between the low-carbon transition (SDG7) and climate adaptation and mitigation. If adequately planned, climate smart infrastructure (SDG9) and novel approaches to traditional sectors (SDG12) improve alignment between the Paris Agreement and the 2030 Agenda for Sustainable Development. Adequate natural resources management is a stable synergy point in LMICs' NDCs: being less resilient to climate impacts, the protection of natural capital fuels the local economy and the safeguard of water (SDG6), biodiversity (SDG15) directly enhances climate adaptation and mitigation.

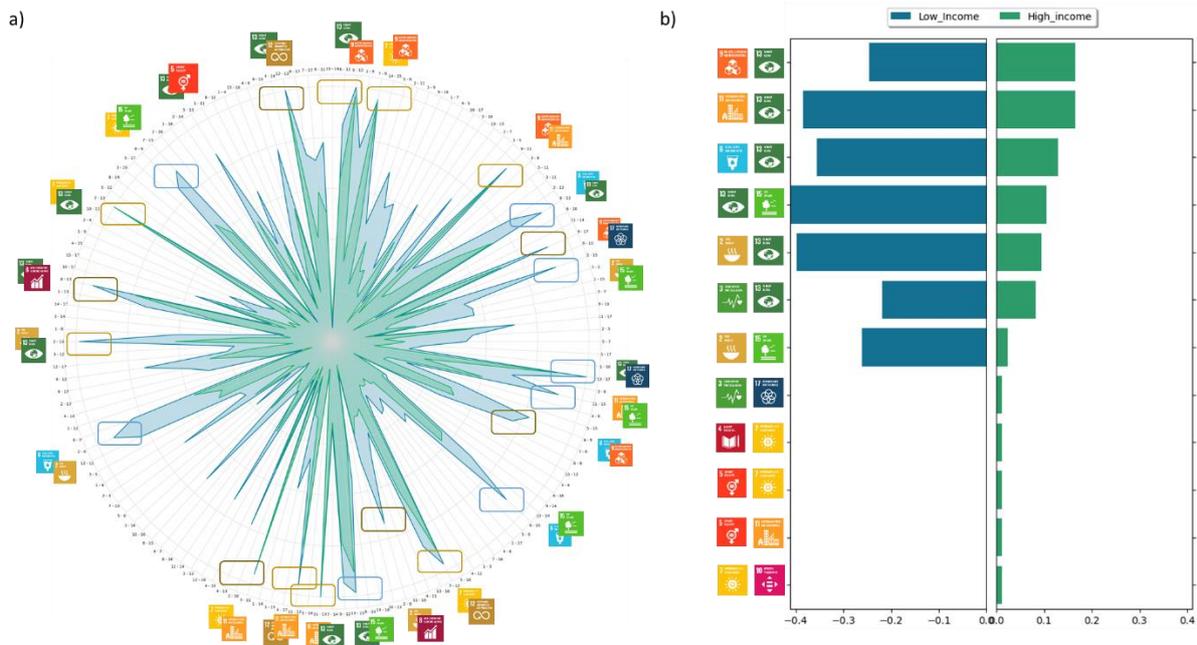

**Figure 5.** Positive (a) and negative (b) stable connections (over space and time) show that high-income and low-income countries converge in tackling synergies and trade-offs, but the magnitude of their pledges differs. a) Lower-income parties discuss positive links with less optimism than their high-income counterparts; b) trade-offs present a scattered landscape with specific connections for each income group. Agriculture is the most represented sector and water management is top mentioned need.

On the trade-offs side (Figure 5b) the NDCs devote considerable time to disaster risk reduction (SDG11) and focus their sectorial needs on transport, especially in sub-Saharan Africa. Water is an issue especially for countries suffering prolonged droughts and poor sanitation. Water critical infrastructures (SDG6) are under threat everywhere in the world due to extreme weather events and a lack of models accounting for physical risks. Sharp differences in covering trade-offs provide an indication of diverging development trajectories in the short and medium run. The energy-infrastructure-community nexus is centered on deploying low-carbon technologies to serve the needs of urban population first. Investments will be concentrated in realizing the transition towards clean energy systems with clear consequences on infrastructures and economic growth.

High-income countries also tackle how service and manufacturing need to change to accommodate the Paris Agreement. Lower-income countries face a "lack of technology" (Bahamas) and "poor technological transfer" (Burundi) especially where infrastructures are not often adequate to climate risks (e.g., Guinea). However, our analysis reveals that persistent topics largely cover the water-energy-food nexus for them (Figure 5b) and question whether climate action shall prevent a lock-in economic effect with the vulnerable becoming even more vulnerable. Agriculture, for example, is a key sector but it contributes to multiple trade-offs with forest management (SDG15) and conflicts for water resources (SDG6). Agriculture-dependent countries may face even bigger obstacles in the presence of frequent climate impacts. Low-income countries recognize that the high dependence on natural resources is a boost for economic growth, but a strong factor risk in the presence of depletion, soil degradation and biodiversity loss (e.g., Malawi). We also find that high-income countries are the ones bringing forward challenges in health (SDG3), education (SDG4), gender equality (SDG5) and inequality (SDG10). Acknowledging that low-income countries tend to neglect these SDGs more than their wealthy counterparts suggests that countries have a hierarchical approach towards the SDGs, preventing a full holistic development from happening.

As six years are left to achieve the 2030 Agenda and the "safe operating space for humanity"[36] is rapidly closing due to anthropogenic factors, systemic approaches and ambitious pledges are urgently needed. Instead of measuring the credibility and ambition degrees of the NDCs as in previous works[37], examining their content sheds light on underlying political choices and leads to significant consequences. Countries differ in their core narratives and center on diverse nexuses. Hence, their development trajectories may diverge in the medium run widening economic inequality and boosting polarization. The NDCs provide a holistic overview of climate actions and pledges and the analysis of interlinkages with other SDGs makes critical contrasts explicit. Despite not containing already committed actions, the automated analysis of NDCs as sustainable development documents suggests three opportunities for improved alignment between climate action and the SDGs.

First, the identification of interlinkages between climate change with other SDGs can improve how international climate and sustainability funds are allocated. In agreement with the GST technical dialogues synthesis report[38], we have found that geographic and sectoral needs are marked but no single recipe exists. An in-depth analysis of the most critical trade-offs can help redesign public economics to better address the actual needs of each country, while also enhancing monitoring efforts. As countries made explicit in their NDCs, insufficient R&D resources in lower-income economies affect infrastructures, especially after natural disasters. Meaningful implementation measures include the involvement of key actors[35] which may connect critically underserved development and geographical areas. This evidence is more urgent than ever as the United Nations has concluded that the world is on track to warm roughly 3.1C[1]. Population dynamics in emerging economies[39] and poorly planned development projects may lead to increased greenhouse emissions, emptying promises and efforts to raise ambitions.

Second, different narratives reveal plausible development trajectories and risks that must be timely detected. As high-income and lower-income countries differ in their core framings of climate action – per quantity, tone and persistency – their priorities affect planning and management activities. To avoid incurring in *lock-in effects*, with countries trapped in climate-affected sectors and busy repairing interest rates, a focus on core synergies and trade-offs can foster a climate-compatible systemic development agenda. Policy changes over time raising questions about revisions and monitoring of plans. AI methods such as our proposed approach allow policy-makers to revisit their strategies whenever needs change. Tools such as LLMs and text-based analyses are also pivotal to collecting meaningful insights when large-scale consultations on these topics happen. The first 2-year long GST process covered 252 hours of meetings and more than 170.000 pages of submitted documents. The new submission rounds may benefit from human-aware and expert-reviewed LLM-extracted content to reveal plausible risks. COP29 reinforced the role of digital technologies and AI in promoting systemic transformation: the COP29 Declaration on Green Digital Action affirms that data-driven novel methodologies support assessment of climate impacts and support accurate standardization. The Declaration further shares the need for policies designed to integrate the "digital and low-emission transition", placing AI at the forefront of a new dynamic interplay between digital, energy and climate goals under the SDGs.

Finally, we find that the NDCs lack sufficient coverage of education (SDG4) and gender (SDG5) dimensions with some exceptions in high-income countries. These pose at-risk efforts to align the 2030 Agenda with the Paris Agreement in two ways. As lower-income countries suggest in some of their submissions, qualified personnel is critically scarce to forecast disaster and to calibrate response making adaptation a priority. In high-income countries, physical climate risks are not adequately incorporated into public policy design and this impacts planning and management of critical infrastructure and urban areas. Gender-responsive policies are also at the heart of the "just transition" debate[40] and lead to positive rebound effects on employment and economic growth[41]. Rather than embracing a hierarchical approach towards the SDGs, prioritizing some over others, we suggest that countries use their NDCs to put forward holistic measures that build on synergies to minimize trade-offs.

The Paris Agreement invites countries to unveil their "long-term strategies" which help countries tackle their near-term plans in the next generation NDCs. The new wave of pledges will cover until 2035. Therefore, a careful and thoughtful assessment of implications for national development policy is crucial to avoid medium-term lock-in effects. With the international community calling for faster and ambitious NDCs implementation, countries will have to promote sector-relevant transformative policies while catalyzing investments from public and private actors. As we presented in this Perspective, an AI-powered, policy-relevant and human-aware analysis can provide the timely insights needed to advance these goals. Moreover, it can highlight if inequalities across income groups and sustainability objectives exist, trying to balance efforts and available resources, as well as informing the sub-national and local policies as well. The current decade is the most critical to meet "the urgency of the moment"[4]: a needs-based agenda will unlock synergies between simultaneous goals and ensure a "just, ordered and responsible"[42] future for all.

**Funding statement**

FL acknowledges funding from the European Union's Horizon Europe research and innovation programme under the Marie Skłodowska-Curie grant agreement No. 101150729. RV, FFN, FL and FM acknowledge the funding provided by Digital Futures, in their Demonstrator-project program. SH acknowledge the grant PID2021-128676OB-I00 funded by MCIN/AEI/10.13039/501100011033 and by "ERDF A Way of Making Europe", by the European Union. JAC is supported by Generalitat Valenciana, Project PROMETEO CIPROM/2022/21.

**Author contributions**

FL: conceptualization, methodology, analysis, writing-original draft, review and editing. FFN, RV: conceptualization, editing, supervision. FM and SH: methodology, analysis, editing. AC, JGM: editing

**Acknowledgements**

The team acknowledges the support and constructive feedback of Oscar Garibo-i-Orts and Lamyae Amhar. The team also acknowledges Prof. Enrica De Cian for her feedback on results. Finally, the team acknowledges the Digital Futures Open Research Days, Digitalize Stockholm and the Euro-Mediteranean Center for offering platforms to discuss and present the results.

**Methods**

The NDCs were downloaded from the UNFCCC NDC Registry (https://unfccc.int/NDCREG) which maintains records of active Parties' submissions in accordance with Article 4 of the Paris Agreement. Only active NDCs were downloaded from the portal (see Supplementary Materials) excluding superseded documents from the analysis. Submissions' metadata were also included (language, status and submission date). 158 countries entered the analysis covering more than 90 percent of the global GHG emissions in per-capita terms (Figure 1M.a) and submitted between 2016 and 2023 (Figure 1M.b). Documents in Arab language and in non-text format (i.e., saved as images) were not included.

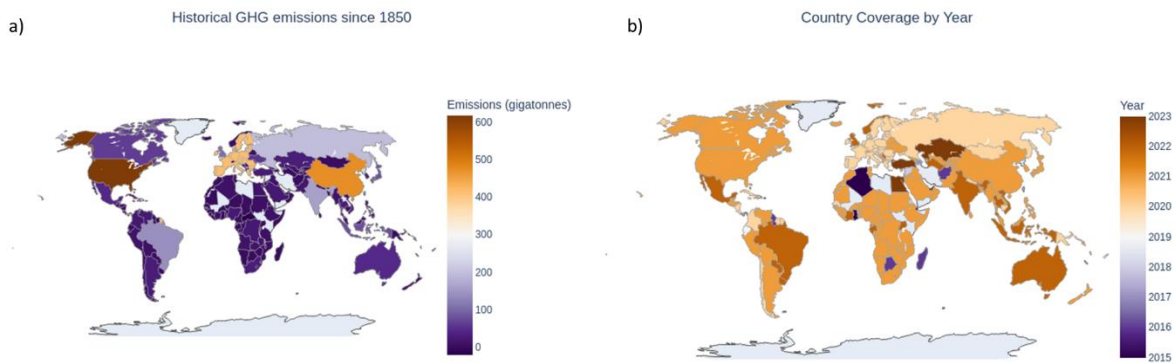

Figure 1M. Countries considered in the analysis per per-capita cumulative emissions (a) and NDC submission year (b)

The NDCs document what countries intend to do to reduce their GHG emissions within a specific timeframe and to mobilize resources to counteract on impacts[1]. Their name ("Nationally Determined Contributions") embody their goal: the NDCs "contribute" to addressing climate change according to a "nationally determined" plan given the Party's circumstances, resources and priorities. This bottom-up process to advance the global effort to limit the temperature rise within 1.5°C is opposed to a top-down vision with internationally-imposed measures[2]. Countries do not follow a standard reporting template for their pledges, but they are left free to describe when they are planning to reach peaking emissions and how they intend to reduce them. This limit reduces affects comparability and transparency[3] as language and style can influence the discourses behind the NDCs. The linguistic assessment of the NDCs have been used to assess how and why countries form coalitions[4] and if ideal clusters based on similar positions shall exist[5]. The style of the NDCs also reveals different responsibilities in terms of emission reduction burden[2] and position countries differently[6] within the international arena. As emphasized in the Paris Agreement itself, "the relationship between climate change actions, responses and impacts with equitable access to sustainable development and eradication of poverty"[1] is undeniable. Therefore, the content of the NDCs reveals important constraining factors and synergic opportunities to advance domestic well-being. Research acknowledged the relationship between sustainable development and climate action in terms of policy coherence[7] and thematic alignment[8]. The SDGs are the preferred lens through which these assessments are put forward. The variety of tools (see Supplementary materials) through which alignment between domestic climate plans (embodied in the NDCs) and the SDGs is assessed demonstrate the wide interest and the policy-relevance of these efforts.

To identify specific connections, the NDCs were downloaded in PDF format and split into self-contained paragraphs in a Python-based routine, Data cleaning was performed to remove *i)* figure and table titles, *ii)* short chunks (less than two words long) or with too many numerical characters (>50 percent), *iii)* sentences below 25 words repeated more than five times within the same documents. Whenever meaningful paragraphs resulted broken into separate lines, a sentence segmentation algorithm was deployed to combine different sentences using a syntactic dependency parsing technique.

To check for the alignment between the NDCs and the SDGs, a two-stage prompt strategy was designed. As the SDGs are not explicitly mentioned in the NDC texts, but implicitly considered throughout the documents, the different stages respond to two sequential, but complementary goals. First, the paragraphs in the NDCs are classified to one, multiple or no SDG. Paragraphs may tackle issues simultaneously when they touch upon co-existing sustainability dimensions. Equally, paragraphs can use a sustainability-neutral language leading to no specific identified SDG. The prompt shall then be flexible enough to allow for variable SDGs per paragraph, without super imposing a pre-determined amount.

Once classified, the second prompt was designed to capture the tone of the identified paragraph with respect to climate adaptation and mitigation. The prompt was structured to assess whether a given text of the NDCs connected to one or more SDGs embodied a positive (assigned 2), neutral (assigned 1) or negative (assigned 0) meaning with respect to wither one or both climate adaptation and mitigation. The purpose of this second step was to grasp what actions and recipes lead to synergies (positive) or trade-offs between the two climate and sustainable development agendas. As the NDCs are forward-looking and programmatic documents, trade-offs were expected in fewer numbers than synergies.

The two-stage prompt was launched using a transformed-based LLM developed by Google Research, Gemini 1.0 Pro[9]. The fast-paced development of LLMs is both an opportunity and a challenge with respect to the optimal and most suitable model. At the time of the analysis, Gemini 1.0 had surpassed and advanced the status quo in large-scale language modeling according to quantitative performance assessments[9]. To further check the suitable of the model and to avoid randomness in the responses, a linguistic-grounded heuristic process assessment was designed. Gemini 1.0 was compared with GPT3.5 in a three-time classification task using three versions of the same prompt (Table 1M). Each version captured a subtle difference in meaning, pushing the model to high level sophistication. Each prompt was applied to the NDCs' paragraphs randomly ordered to check for variability in responses. *Ceteris paribus*, Gemini 1.0 proved stable within the rounds and across the different versions.

**Table 1M. Alternative prompts**

| | |
|---|---|
| Version #1: quantitative assessment | "Assign the following text to the top three SDGs based on their **dominance**" |
| Version #2: value assessment | "Assign the following text to the top three SDGs per **relevance**" |
| Version #3: quali-quantity mixed | "Assign the following text to the top three SDGs based on their **prominence**" |

The two prompts were designed following an iterative trial-and-error method with a climate policy expert and a programmer translating the goals to Python-based language. Seventeen versions of the two-stage prompt (see Supplementary Materials) were tested and outputs manually screened to assess reliability. As Gemini 1.0 already has prior knowledge about the SDGs, no specific text or context around them was provided in the first stage, limiting arbitrary choices. Stage one of the strategy reads:

*"Assign the following text to all relevant SDGs (strictly from the following list: 1) No poverty 2) Zero Hunger 3) Good health and well-being 4) Quality education 5) Gender equality 6) Clean Water and Sanitation 7) Affordable and clean energy 8) Decent work and economic growth 9) Industry, innovation and infrastructure 10) Reduced inequalities 11) Sustainable cities and communities 12) Responsible consumption and production 13) Climate action 14) Life below water 15) Life on land 16) Peace, justice and strong institutions 17) Partnerships for the goals). If a paragraph tackles non relevant issues with respect to any SDG, assign 0. An example of a correct output is -->*
*SDG (pertinent number): Name of SDG \n*
*Reason SDG (pertinent number): clear justification \n*
*SDG (pertinent number): Name of SDG \n*
*Reason SDG (pertinent number): clear justification ... and so on with as many pertinent SDGs."*

Stage-two prompt uses the output of stage one and reads as following:

*"Use the following rules to interpret a paragraph: Consider climate adaptation as the adjustment in natural or human systems in response to actual or expected climatic stimuli or their effects, which moderates harm or exploits beneficial opportunities. Also consider, climate mitigation as an anthropogenic intervention to reduce the sources or enhance the sinks of greenhouse gas. Assign to each paragraph one and one only number between 0, 1 and 2. Assign 0 if the paragraph explains or present an action or a set of actions which pose concrete risks to at least one between climate adaptation and mitigation; assign 1 if the paragraph is neutral with respect to climate adaptation and mitigation and does not express or discuss any concrete opportunity or risk for the country; assign 2 if the paragraph explains or present an action or a set of actions which pose concrete opportunities to at least one between climate adaptation and mitigation".*

The output of the AI-based routine is a dense network of interactions with SDGs as nodes linked to one another in presence of at least one paragraph shared. Nodes' attributes respond to the SDG classification by Norström et al.[10] with clusters responding to economy, society and environment. Links' attributes relate to the type of node (positive, neutral or negative claim with respect to climate adaptation and mitigation). Weights are represented by the frequency of mention. Connections are represented in the form a heatmap (Figure 1c): weights are computed as bilateral connections over the raw-level maximum pairwise link. This approach overcomes the statistical bias introduced by long documents with highly mentioned connections as NDCs are all treated equally. In the global landscape, connection strengths range from a minimum of 0.04 (SDG7-SDG4) to a maximum of 1 (as for SDG13 which connects to every other SDG by construction). However, countries are not equal: they have heterogeneous socio-economic and technical characteristics which affect priorities, development challenges and ultimately pathways. Income classes as derived from the World Bank Group are used to position countries into meaningful clusters. The classification is built on previous year's GNI per capita and allocates countries in low, lower-middle, upper-middle and high income groups. While not perfectly correlated, per-capita income classes well reflect the evolution of per-capita emissions and can then be used to describe the climate ambition need to improve country's emission profiles. Higher (high and upper-middle) income countries are more distributed across emitter classes than lower (low and lower-middle) income countries (Figure 2M.a). This is also consistent with differences in SDG-SDG connection between income and emission classes. The sum of square difference between the edges of each SDG-SDG averaged connection graph, reveals that low-income connections are highly correlated with low-emitter ones and that higher income links are correlated to both middle and high emitter's (Figure 2M.b).

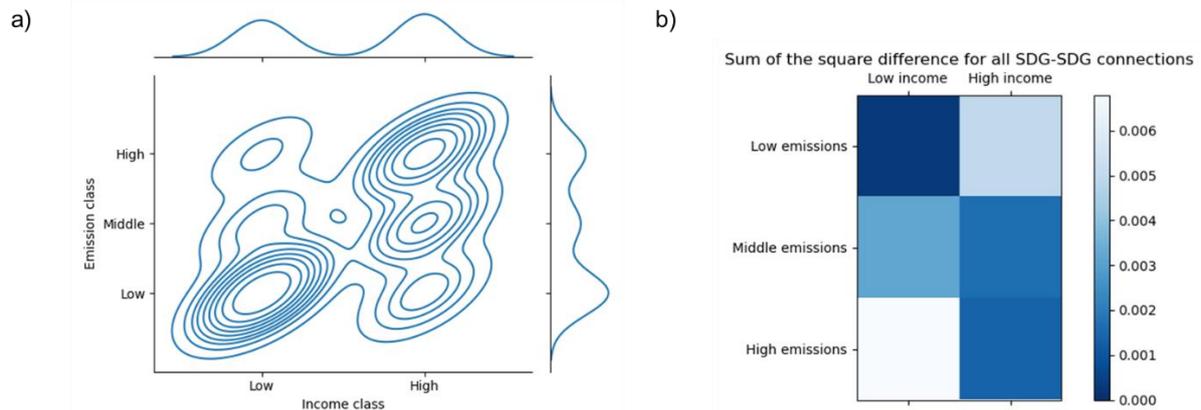

Figure 2M. Correlation between per-capita emission and per-capita income classes in absolute terms (a) and in graph-term (b)

Class-relevant interlinkages graphs are presented to compare and contrast how countries tackle the climate-sustainability link. However, some connections remain stable across the whole domain (persistent links, Figure 4) and represent the backbone of the synergy and trade-offs opportunities. To identify these stable or persistent connections, we assign a value of 1 for each pair of SDGs in a country's NDC contribution if that connection is present, and a 0 if it is not. These values are then summed across all countries for each pair of SDGs, and the sum is divided by the total number of countries to compute the persistence metric. This method minimizes bias introduced by individual countries that may disproportionately focus on specific connections, providing a clearer picture of common interlinkages.

# Supplementary materials

**Table A. Glossary**

| Term | Definition | Source |
|------|-----------|--------|
| Climate action | Set of policies – at local and global level – aimed at mitigating hazardous effects of climate change. In this context, we call "climate action" the set of pledges, proposals, projects and programs which advance both climate adaptation and mitigation. | |
| Nationally Determined Contribution (NDCs) | "A climate action plan to cut emissions and adapt to climate impacts. Each Party to the Paris Agreement is required to establish an NDC and update it every five years" | UNFCCC NDC: https://unfccc.int/process-and-meetings/the-paris-agreement/nationally-determined-contributions-ndcs |
| First, updated and second submission of the NDCs | Since the launch of the Paris agreement, countries have engaged in two submission rounds. Countries interchangly distinguish between "first" and "second" or "first and "enhanced" version. In this article, we use the terminology "first" and "second" submission to distinguish between two different rounds. | UNFCCC NDC Registry: https://unfccc.int/NDCREG |
| Large Language Model | An AI systems which uses language as input and training material. An AI system is "a computational representation that encompasses processes, objects, ideas, people and interactions. Language models vary in language and size. AI language models are often characterised by their parameter count and layers and accuracy" | pp.22, OECD (2023)[1] |
| Natural language Processing (NLP) | "Computer programs and tools that automate natural language functions by analysing, producing, modifying, or responding to human texts and speech" | pp.14, OECD (2023)[1] |

**Table B. List of countries and respective classification**

| Year | Country | Income group | Geography | Emission group |
|------|---------|--------------|-----------|----------------|
| 2016 | Afghanistan | Low income | South Asia | Low emissions |
| 2021 | Albania | Upper middle income | Europe & Central Asia | Low emissions |
| 2015 | Algeria | Lower middle income | Middle East & North Africa | Middle emissions |
| 2022 | Andorra | High income | Europe & Central Asia | |
| 2021 | Angola | Lower middle income | Sub-Saharan Africa | Low emissions |
| 2021 | Antigua and Barbuda | High income | Latin America & Caribbean | |
| 2021 | Argentina | Upper middle income | Latin America & Caribbean | High emissions |
| 2021 | Armenia | Upper middle income | Europe & Central Asia | Low emissions |
| 2022 | Australia | High income | East Asia & Pacific | High emissions |
| 2017 | Azerbaijan | Upper middle income | Europe & Central Asia | Middle emissions |
| 2022 | Bahamas | High income | Latin America & Caribbean | High emissions |
| 2021 | Bahrain | High income | Middle East & North Africa | High emissions |
| 2021 | Bangladesh | Lower middle income | South Asia | Low emissions |
| 2021 | Barbados | High income | Latin America & Caribbean | Middle emissions |
| 2021 | Belarus | Upper middle income | Europe & Central Asia | Middle emissions |
| 2021 | Belize | Upper middle income | Latin America & Caribbean | High emissions |
| 2021 | Benin | Lower middle income | Sub-Saharan Africa | Low emissions |
| 2021 | Bhutan | Lower middle income | South Asia | Low emissions |
| 2022 | Bolivia | Lower middle income | Latin America & Caribbean | High emissions |
| 2021 | Bosnia and Herzegovina | Upper middle income | Europe & Central Asia | Middle emissions |
| 2016 | Botswana | Upper middle income | Sub-Saharan Africa | High emissions |
| 2022 | Brazil | Upper middle income | Latin America & Caribbean | High emissions |
| 2020 | Brunei Darussalam | High income | East Asia & Pacific | High emissions |
| 2021 | Burkina Faso | Low income | Sub-Saharan Africa | Low emissions |
| 2020 | Burundi | Low income | Sub-Saharan Africa | Low emissions |
| 2021 | Cabo Verde | Lower middle income | Sub-Saharan Africa | Low emissions |
| 2020 | Cambodia | Lower middle income | East Asia & Pacific | Low emissions |
| 2021 | Cameroon | Lower middle income | Sub-Saharan Africa | Low emissions |
| 2021 | Canada | High income | North America | High emissions |

| 2021 | Central African Republic | Low income | Sub-Saharan Africa | Low emissions |
|---|---|---|---|---|
| 2021 | Chad | Low income | Sub-Saharan Africa | Middle emissions |
| 2020 | Chile | High income | Latin America & Caribbean | Low emissions |
| 2021 | China | Upper middle income | East Asia & Pacific | High emissions |
| 2020 | Colombia | Upper middle income | Latin America & Caribbean | Middle emissions |
| 2021 | Comoros | Lower middle income | Sub-Saharan Africa | Low emissions |
| 2021 | Congo | Lower middle income | Sub-Saharan Africa | Middle emissions |
| 2021 | Congo, Dem. Rep. | Low income | Sub-Saharan Africa | High emissions |
| 2016 | Cook Island | High income | East Asia & Pacific | |
| 2020 | Costa Rica | Upper middle income | Latin America & Caribbean | Low emissions |
| 2022 | Côte d'Ivoire | Lower middle income | Sub-Saharan Africa | Low emissions |
| 2020 | Cuba | Upper middle income | Latin America & Caribbean | Low emissions |
| 2015 | Djibouti | Lower middle income | Middle East & North Africa | Low emissions |
| 2022 | Dominica | Upper middle income | Latin America & Caribbean | Low emissions |
| 2019 | Ecuador | Upper middle income | Latin America & Caribbean | Middle emissions |
| 2023 | Egypt | Lower middle income | Middle East & North Africa | Low emissions |
| 2021 | El Salvador | Lower middle income | Latin America & Caribbean | Low emissions |
| 2022 | Equatorial Guinea | Upper middle income | Sub-Saharan Africa | High emissions |
| 2018 | Eritrea | Low income | Sub-Saharan Africa | Low emissions |
| 2021 | Eswatini | Lower middle income | Sub-Saharan Africa | Low emissions |
| 2021 | Ethiopia | Low income | Sub-Saharan Africa | Low emissions |
| 2020 | European Union | High income | Europe & Central Asia | High emissions |
| 2020 | Fiji | Upper middle income | East Asia & Pacific | Low emissions |
| 2022 | Gabon | Upper middle income | Sub-Saharan Africa | High emissions |
| 2021 | Gambia | Low income | Sub-Saharan Africa | Low emissions |
| 2021 | Georgia | Upper middle income | Europe & Central Asia | Middle emissions |
| 2015 | Ghana | Lower middle income | Sub-Saharan Africa | Low emissions |
| 2020 | Grenada | Upper middle income | Latin America & Caribbean | |
| 2021 | Guatemala | Upper middle income | Latin America & Caribbean | Low emissions |
| 2021 | Guinea | Low income | Sub-Saharan Africa | Low emissions |

| 2021 | Guinea-Bissau | Low income | Sub-Saharan Africa | Middle emissions |
|------|---------------|------------|---------------------|------------------|
| 2016 | Guyana | Upper middle income | Latin America & Caribbean | High emissions |
| 2021 | Haiti | Low income | Latin America & Caribbean | Low emissions |
| 2021 | Honduras | Lower middle income | Latin America & Caribbean | Low emissions |
| 2021 | Iceland | High income | Europe & Central Asia | High emissions |
| 2022 | India | Lower middle income | South Asia | Low emissions |
| 2022 | Indonesia | Upper middle income | East Asia & Pacific | Middle emissions |
| 2021 | Israel | High income | Middle East & North Africa | Middle emissions |
| 2020 | Jamaica | Upper middle income | Latin America & Caribbean | Middle emissions |
| 2021 | Japan | High income | East Asia & Pacific | High emissions |
| 2021 | Jordan | Upper middle income | Middle East & North Africa | Low emissions |
| 2023 | Kazakhstan | Upper middle income | Europe & Central Asia | High emissions |
| 2021 | Korea Republic | High income | East Asia & Pacific | High emissions |
| 2021 | Kuwait | High income | Middle East & North Africa | High emissions |
| 2021 | Kyrgyz Republic | Lower middle income | Europe & Central Asia | Low emissions |
| 2021 | Lao PDR | Lower middle income | East Asia & Pacific | |
| 2020 | Lebanon | Upper middle income | Middle East & North Africa | Middle emissions |
| 2017 | Lesotho | Lower middle income | Sub-Saharan Africa | Low emissions |
| 2021 | Liberia | Low income | Sub-Saharan Africa | Low emissions |
| 2017 | Liechtenstein | High income | Europe & Central Asia | |
| 2016 | Madagascar | Low income | Sub-Saharan Africa | Low emissions |
| 2021 | Malawi | Low income | Sub-Saharan Africa | Low emissions |
| 2021 | Malaysia | Upper middle income | East Asia & Pacific | High emissions |
| 2020 | Maldives | Upper middle income | South Asia | Middle emissions |
| 2021 | Mauritania | Lower middle income | Sub-Saharan Africa | Middle emissions |
| 2021 | Mauritius | High income | Sub-Saharan Africa | Middle emissions |
| 2022 | Mexico | Upper middle income | Latin America & Caribbean | Middle emissions |
| 2022 | Micronesia | Lower middle income | East Asia & Pacific | |
| 2020 | Moldova | Lower middle income | Europe & Central Asia | Low emissions |
| 2020 | Monaco | High income | Europe & Central Asia | |

| 2020 | Mongolia | Lower middle income | East Asia & Pacific | High emissions |
|---|---|---|---|---|
| 2021 | Montenegro | Upper middle income | Europe & Central Asia | Middle emissions |
| 2021 | Morocco | Lower middle income | Middle East & North Africa | Low emissions |
| 2021 | Mozambique | Low income | Sub-Saharan Africa | Low emissions |
| 2021 | Myanmar | Lower middle income | East Asia & Pacific | Middle emissions |
| 2021 | Namibia | Upper middle income | Sub-Saharan Africa | Middle emissions |
| 2021 | Nauru | High income | East Asia & Pacific | |
| 2020 | Nepal | Lower middle income | South Asia | Low emissions |
| 2021 | New Zealand | High income | East Asia & Pacific | High emissions |
| 2020 | Nicaragua | Lower middle income | Latin America & Caribbean | Middle emissions |
| 2021 | Niger | Low income | Sub-Saharan Africa | Low emissions |
| 2021 | Nigeria | Lower middle income | Sub-Saharan Africa | Low emissions |
| 2016 | Niue | High income | East Asia & Pacific | |
| 2021 | North Macedonia | Upper middle income | Europe & Central Asia | High emissions |
| 2022 | Norway | High income | Europe & Central Asia | High emissions |
| 2021 | Oman | High income | Middle East & North Africa | High emissions |
| 2021 | Pakistan | Lower middle income | South Asia | Low emissions |
| 2015 | Palau | High income | East Asia & Pacific | |
| 2020 | Papua New Guinea | Lower middle income | East Asia & Pacific | High emissions |
| 2022 | Paraguay | Upper middle income | Latin America & Caribbean | High emissions |
| 2021 | Peru | Upper middle income | Latin America & Caribbean | Middle emissions |
| 2021 | Philippines | Lower middle income | East Asia & Pacific | Low emissions |
| 2020 | Russian Federation | Upper middle income | Europe & Central Asia | High emissions |
| 2020 | Rwanda | Low income | Sub-Saharan Africa | Low emissions |
| 2021 | Saint Lucia | Upper middle income | Latin America & Caribbean | Low emissions |
| 2021 | Samoa | Upper middle income | East Asia & Pacific | Low emissions |
| 2015 | San Marino | High income | Europe & Central Asia | |
| 2021 | São Tomé and Principe | Lower middle income | Sub-Saharan Africa | Low emissions |
| 2021 | Saudi Arabia | High income | Middle East & North Africa | High emissions |
| 2020 | Senegal | Lower middle income | Sub-Saharan Africa | Low emissions |

| 2022 | Serbia | Upper middle income | Europe & Central Asia | Middle emissions |
|------|--------|---------------------|----------------------|------------------|
| 2021 | Seychelles | High income | Sub-Saharan Africa | |
| 2021 | Sierra Leone | Low income | Sub-Saharan Africa | Low emissions |
| 2022 | Singapore | High income | East Asia & Pacific | High emissions |
| 2021 | Solomon Islands | Lower middle income | East Asia & Pacific | High emissions |
| 2021 | Somalia | Low income | Sub-Saharan Africa | Low emissions |
| 2021 | South Africa | Upper middle income | Sub-Saharan Africa | Middle emissions |
| 2021 | Sri Lanka | Lower middle income | South Asia | Low emissions |
| 2021 | St. Kitts and Nevis | High income | Latin America & Caribbean | |
| 2015 | St. Vincent and the Grenadines | Upper middle income | Latin America & Caribbean | Low emissions |
| 2021 | State of Palestine | Lower middle income | Middle East & North Africa | |
| 2021 | Sudan | Low income | Sub-Saharan Africa | Low emissions |
| 2020 | Suriname | Upper middle income | Latin America & Caribbean | High emissions |
| 2021 | Switzerland | High income | Europe & Central Asia | Middle emissions |
| 2018 | Syrian Arab Republic | Low income | Middle East & North Africa | |
| 2021 | Tajikistan | Low income | Europe & Central Asia | Low emissions |
| 2021 | Tanzania | Lower middle income | Sub-Saharan Africa | Low emissions |
| 2022 | Thailand | Upper middle income | East Asia & Pacific | Middle emissions |
| 2022 | Timor-Leste | Lower middle income | East Asia & Pacific | High emissions |
| 2021 | Togo | Low income | Sub-Saharan Africa | Low emissions |
| 2020 | Tonga | Upper middle income | East Asia & Pacific | Low emissions |
| 2018 | Trinidad and Tobago | High income | Latin America & Caribbean | High emissions |
| 2021 | Tunisia | Lower middle income | Middle East & North Africa | Low emissions |
| 2023 | Turkey | Upper middle income | Europe & Central Asia | Middle emissions |
| 2022 | Turkmenistan | Upper middle income | Europe & Central Asia | High emissions |
| 2022 | Tuvalu | Upper middle income | East Asia & Pacific | |
| 2022 | Uganda | Low income | Sub-Saharan Africa | Low emissions |
| 2021 | Ukraine | Lower middle income | Europe & Central Asia | Low emissions |
| 2023 | United Arab Emirates | High income | Middle East & North Africa | High emissions |
| 2022 | United Kingdom | High income | Europe & Central Asia | Middle emissions |

| 2021 | United States | High income | North America | High emissions |
|------|---------------|-------------|---------------|----------------|
| 2022 | Uruguay | High income | Latin America & Caribbean | High emissions |
| 2021 | Uzbekistan | Lower middle income | Europe & Central Asia | Middle emissions |
| 2021 | Vanuatu | Lower middle income | East Asia & Pacific | Low emissions |
| 2023 | Vatican City State | High income | Europé & Central Asia | |
| 2021 | Venezuela | Upper middle income | Latin America & Caribbean | Middle emissions |
| 2022 | Vietnam | Lower middle income | East Asia & Pacific | Middle emissions |
| 2021 | Zambia | Lower middle income | Sub-Saharan Africa | Low emissions |
| 2021 | Zimbabwe | Lower middle income | Sub-Saharan Africa | Middle emissions |

**Table C. Tools to assess NDC-SDGs alignment**

| Name | Link | Management by | Year | Description |
|------|------|---------------|------|-------------|
| NDC-SDG Linkages | https://www.climatewatchdata.org/ndcs-sdg | ClimateWatch | Online, updated up to May 2021 | Identify potential alignment between the targets, actions, policy measures and needs in countries' Nationally Determined Contributions (NDCs) and the targets of the Sustainable Development Goals (SDGs). |
| NDC-SDG Connections | https://klimalog.idos-research.de/ndc-sdg/ | German Institute of Development and Sustainability and Stockholm Environmental Institute | Online, updated | NDC-SDG Connections is a joint initiative of the German Institute of Development and Sustainability (IDOS) and the Stockholm Environment Institute (SEI). The research and visualisation project aims at illuminating synergies between the 2030 Agenda for Sustainable Development and the Paris Agreement, and at identifying entry points for coherent policies that promote just, sustainable and climate-smart development. |
| Examining the Alignment between the Intended Nationally Determined Contributions and Sustainable | https://www.wri.org/research/examining-alignment-between-intended-nationally-determined-contributions-and-sustainable | World Resources Institute | 2016 | The paper explores the extent of alignment between the climate and the sustainable development agendas demonstrating that climate actions communicated in the Intended Nationally Determined Contributions under the Paris Agreement have the potential to generate mutual benefits with at least 154 of the 169 SDG targets. |
| Development Goals | | | | |
| Synergy Solutions for a World in Crisis: Tackling Climate and SDG Action Together | https://sdgs.un.org/synergy-solutions-world-crisis-tackling-climate-and-sdg-action-together | UNDESA | 2023 | A report by a group of independent experts outlining steps governments should take to maximize the impact of policies and actions by tackling the climate and sustainable development crises at the same time, creating synergies. |

| Year | Country | Country_Code | Income_class | Income | Vulnerability_CC | Readiness | VUL | Geo | Region | Emission_class |
|---|---|---|---|---|---|---|---|---|---|---|
| 2016 | Afghanistan | AFG | Low income | 1 | 0,59 | 0,24 | 1 | South Asia | 7 | Low emissions |
| 2021 | Albania | ALB | Upper middle income | 3 | 0,39 | 0,41 | 3 | Europe & Central Asia | 3 | Low emissions |
| 2015 | Algeria | DZA | Lower middle income | 2 | 0,37 | 0,33 | 2 | Middle East & North Africa | 1 | Middle emissions |
| 2022 | Andorra | AND | High income | 4 | | 0,47 | 3 | Europe & Central Asia | 3 | |
| 2021 | Angola | AGO | Lower middle income | 2 | 0,5 | 0,26 | 1 | Sub-Saharan Africa | 5 | Low emissions |
| 2021 | Antigua and Barbuda | ATG | High income | 4 | 0,46 | 0,44 | 4 | Latin America & Caribbean | 4 | |
| 2021 | Argentina | ARG | Upper middle income | 3 | 0,38 | 0,37 | 2 | Latin America & Caribbean | 4 | High emissions |
| 2021 | Armenia | ARM | Upper middle income | 3 | 0,36 | 0,5 | 3 | Europe & Central Asia | 3 | Low emissions |

| 2022 | Australia | AUS | High income | 4 | 0,31 | 0,69 | 3 | East Asia & Pacific | 6 | High emissions |
|---|---|---|---|---|---|---|---|---|---|---|
| 2017 | Azerbaijan | AZE | Upper middle income | 3 | 0,38 | 0,44 | 3 | Europe & Central Asia | 3 | Middle emissions |
| 2022 | Bahamas | BHS | High income | 4 | 0,45 | 0,42 | 4 | Latin America & Caribbean | 4 | High emissions |
| 2021 | Bahrain | BHR | High income | 4 | 0,44 | 0,51 | 4 | Middle East & North Africa | 1 | High emissions |
| 2021 | Bangladesh | BGD | Lower middle income | 2 | 0,53 | 0,28 | 1 | South Asia | 7 | Low emissions |
| 2021 | Barbados | BRB | High income | 4 | 0,38 | 0,53 | 3 | Latin America & Caribbean | 4 | Middle emissions |
| 2021 | Belarus | BLR | Upper middle income | 3 | 0,33 | 0,49 | 3 | Europe & Central Asia | 3 | Middle emissions |
| 2021 | Belize | BLZ | Upper middle income | 3 | 0,45 | 0,33 | 1 | Latin America & Caribbean | 4 | High emissions |
| 2021 | Benin | BEN | Lower middle income | 2 | 0,55 | 0,33 | 1 | Sub-Saharan Africa | 5 | Low emissions |
| 2021 | Benin | BEN | Lower middle income | 2 | 0,55 | 0,33 | 1 | Sub-Saharan Africa | 5 | Low emissions |
| 2021 | Bhutan | BTN | Lower middle income | 2 | 0,51 | 0,48 | 4 | South Asia | 7 | Low emissions |
| 20 21 | Bolivia | BOL | Lower middle income | 2 | 0,44 | 0,28 | 1 | Latin America & Caribbean | 4 | High emissions |
| 20 21 | Bosnia and Herzegovina | BIH | Upper middle income | 3 | 0,34 | 0,36 | 2 | Europe & Central Asia | 3 | Middle emissions |
| 20 16 | Botswana | BWA | Upper middle income | 3 | 0,41 | 0,43 | 3 | Sub-Saharan Africa | 5 | High emissions |
| 20 22 | Brazil | BRA | Upper middle income | 3 | 0,37 | 0,35 | 2 | Latin America & Caribbean | 4 | High emissions |
| 20 20 | Brunei Darussalam | BRN | High income | 4 | 0,39 | 0,53 | 3 | East Asia & Pacific | 6 | High emissions |
| 20 21 | Burkina Faso | BFA | Low income | 1 | 0,53 | 0,28 | 1 | Sub-Saharan Africa | 5 | Low emissions |
| 20 20 | Burundi | BDI | Low income | 1 | 0,55 | 0,26 | 1 | Sub-Saharan Africa | 5 | Low emissions |
| 20 21 | Cabo Verde | CPV | Lower middle income | 2 | 0,42 | 0,45 | 3 | Sub-Saharan Africa | 5 | Low emissions |
| 20 20 | Cambodia | KHM | Lower middle income | 2 | 0,48 | 0,28 | 1 | East Asia & Pacific | 6 | Low emissions |
| 20 21 | Cameroon | CMR | Lower middle income | 2 | 0,46 | 0,26 | 1 | Sub-Saharan Africa | 5 | Low emissions |
| 20 21 | Canada | CAN | High income | 4 | 0,28 | 0,64 | 3 | North America | | High emissions |
| 20 21 | Central African Republic | CAF | Low income | 1 | 0,58 | 0,13 | 1 | Sub-Saharan Africa | 5 | Low emissi ons |
| 20 21 | Chad | TCD | Low income | 1 | 0,65 | 0,19 | 1 | Sub-Saharan Africa | 5 | Middle emissi ons |
| 20 20 | Chile | CHL | High income | 4 | 0,32 | 0,53 | 3 | Latin America & Caribbean | 4 | Low emissi ons |
| 20 21 | China | CHN | Upper middle income | 3 | 0,38 | 0,55 | 3 | East Asia & Pacific | 6 | High emissi ons |
| 20 20 | Colombia | COL | Upper middle income | 3 | 0,41 | 0,37 | 2 | Latin America & Caribbean | 4 | Middle emissi ons |
| 20 20 | Colombia | COL | Upper middle income | 3 | 0,41 | 0,37 | 2 | Latin America & Caribbean | 4 | Middle emissi ons |
| 20 20 | Colombia | COL | Upper middle income | 3 | 0,41 | 0,37 | 2 | Latin America & Caribbean | 4 | Middle emissi ons |
| 20 20 | Colombia | COL | Upper middle income | 3 | 0,41 | 0,37 | 2 | Latin America & Caribbean | 4 | Middle emissi ons |
| 20 20 | Colombia | COL | Upper middle income | 3 | 0,41 | 0,37 | 2 | Latin America & Caribbean | 4 | Middle emissi ons |
| 20 20 | Colombia | COL | Upper middle income | 3 | 0,41 | 0,37 | 2 | Latin America & Caribbean | 4 | Middle emissi ons |
| 20 20 | Colombia | COL | Upper middle income | 3 | 0,41 | 0,37 | 2 | Latin America & Caribbean | 4 | Middle emissi ons |

| 2020 | Colombia | COL | Upper middle income | 3 | 0,41 | 0,37 | 2 | Latin America & Caribbean | 4 | Middle emissions |
|------|----------|-----|---------------------|---|------|------|---|---------------------------|---|------------------|
| 2020 | Colombia | COL | Upper middle income | 3 | 0,41 | 0,37 | 2 | Latin America & Caribbean | 4 | Middle emissions |
| 2020 | Colombia | COL | Upper middle income | 3 | 0,41 | 0,37 | 2 | Latin America & Caribbean | 4 | Middle emissions |
| 2020 | Colombia | COL | Upper middle income | 3 | 0,41 | 0,37 | 2 | Latin America & Caribbean | 4 | Middle emissions |
| 2020 | Colombia | COL | Upper middle income | 3 | 0,41 | 0,37 | 2 | Latin America & Caribbean | 4 | Middle emissions |
| 2020 | Colombia | COL | Upper middle income | 3 | 0,41 | 0,37 | 2 | Latin America & Caribbean | 4 | Middle emissions |
| 2020 | Colombia | COL | Upper middle income | 3 | 0,41 | 0,37 | 2 | Latin America & Caribbean | 4 | Middle emissions |
| 2020 | Colombia | COL | Upper middle income | 3 | 0,41 | 0,37 | 2 | Latin America & Caribbean | 4 | Middle emissions |
| 2020 | Colombia | COL | Upper middle income | 3 | 0,41 | 0,37 | 2 | Latin America & Caribbean | 4 | Middle emissions |
| 2020 | Colombia | COL | Upper middle income | 3 | 0,41 | 0,37 | 2 | Latin America & Caribbean | 4 | Middle emissions |
| 2021 | Comoros | COM | Lower middle income | 2 | 0,52 | 0,28 | 1 | Sub-Saharan Africa | 5 | Low emissions |

| 20 21 | Congo | COG | Lower middle income | 2 | 0,52 | 0,22 | 1 | Sub-Saharan Africa | 5 | Middle emissions |
|---|---|---|---|---|---|---|---|---|---|---|
| 20 21 | Congo, Dem. Rep. | COD | Low income | 1 | 0,56 | 0,21 | 1 | Sub-Saharan Africa | 5 | High emissions |
| 20 16 | Cook Island | COK | High income | 4 | | | 2 | East Asia & Pacific | 6 | |
| 20 20 | Costa Rica | CRI | Upper middle income | 3 | 0,37 | 0,45 | 3 | Latin America & Caribbean | 4 | Low emissions |
| 20 22 | Côte d'Ivoire | CIV | Lower middle income | 2 | 0,48 | 0,3 | 1 | Sub-Saharan Africa | 5 | Low emissions |
| 20 20 | Cuba | CUB | Upper middle income | 3 | 0,42 | 0,35 | 2 | Latin America & Caribbean | 4 | Low emissions |
| 20 15 | Djibouti | DJI | Lower middle income | 2 | 0,47 | 0,32 | 1 | Middle East & North Africa | 1 | Low emissions |
| 20 22 | Dominica | DMA | Upper middle income | 3 | 0,45 | 0,52 | 4 | Latin America & Caribbean | 4 | Low emissions |
| 20 19 | Ecuador | ECU | Upper middle income | 3 | 0,45 | 0,34 | 1 | Latin America & Caribbean | 4 | Middle emissions |
| 20 23 | Egypt | EGY | Lower middle income | 2 | 0,42 | 0,35 | 2 | Middle East & North Africa | 1 | Low emissions |
| 20 21 | El Salvador | SLV | Lower middle income | 2 | 0,42 | 0,33 | 2 | Latin America & Caribbean | 4 | Low emissions |
| 20 22 | Equatorial Guinea | GNQ | Upper middle income | 3 | 0,41 | 0,24 | 2 | Sub-Saharan Africa | 5 | High emissions |
| 20 18 | Eritrea | ERI | Low income | 1 | 0,6 | 0,22 | 1 | Sub-Saharan Africa | 5 | Low emissi ons |
| 20 21 | Eswatini | SWZ | Lower middle income | 2 | 0,47 | 0,31 | 1 | Sub-Saharan Africa | 5 | Low emissi ons |
| 20 21 | Ethiopia | ETH | Low income | 1 | 0,54 | 0,29 | 1 | Sub-Saharan Africa | 5 | Low emissi ons |
| 20 20 | European Union | EU | High income | 4 | | | 2 | Europe & Central Asia | 3 | High emissi ons |
| 20 20 | Fiji | FJI | Upper middle income | 3 | 0,45 | 0,47 | 4 | East Asia & Pacific | 6 | Low emissi ons |
| 20 22 | Gabon | GAB | Upper middle income | 3 | 0,44 | 0,3 | 1 | Sub-Saharan Africa | 5 | High emissi ons |
| 20 21 | Gambia | GMB | Low income | 1 | 0,52 | 0,32 | 1 | Sub-Saharan Africa | 5 | Low emissi ons |
| 20 21 | Georgia | GEO | Upper middle income | 3 | 0,39 | 0,56 | 3 | Europe & Central Asia | 3 | Middle emissi ons |
| 20 15 | Ghana | GHA | Lower middle income | 2 | 0,44 | 0,34 | 1 | Sub-Saharan Africa | 5 | Low emissi ons |
| 20 20 | Grenada | GRD | Upper middle income | 3 | 0,38 | 0,47 | 3 | Latin America & Caribbean | 4 | |
| 20 21 | Guatemala | GTM | Upper middle income | 3 | 0,43 | 0,31 | 1 | Latin America & Caribbean | 4 | Low emissi ons |
| 20 21 | Guinea | GIN | Low income | 1 | 0,54 | 0,3 | 1 | Sub-Saharan Africa | 5 | Low emissi ons |
| 2021 | Guinea-Bissau | GNB | Low income | 1 | 0,62 | 0,27 | 1 | Sub-Saharan Africa | 5 | Middle emissions |
| 2016 | Guyana | GUY | Upper middle income | 3 | 0,42 | 0,32 | 2 | Latin America & Caribbean | 4 | High emissions |
| 2021 | Haiti | HTI | Low income | 1 | 0,51 | 0,22 | 1 | Latin America & Caribbean | 4 | Low emissions |
| 2021 | Honduras | HND | Lower middle income | 2 | 0,45 | 0,25 | 1 | Latin America & Caribbean | 4 | Low emissions |
| 2021 | Iceland | ISL | High income | 4 | 0,3 | 0,72 | 3 | Europe & Central Asia | 3 | High emissions |
| 2022 | India | IND | Lower middle income | 2 | 0,49 | 0,38 | 1 | South Asia | 7 | Low emissions |
| 2022 | Indonesia | IDN | Upper middle income | 3 | 0,44 | 0,39 | 4 | East Asia & Pacific | 6 | Middle emissions |
| 2021 | Iraq | IRQ | Upper middle income | 3 | 0,44 | 0,3 | 1 | Middle East & North Africa | 1 | Middle emissions |
| 2021 | Israel | ISR | High income | 4 | 0,3 | 0,54 | 3 | Middle East & North Africa | 1 | Middle emissions |
| 2020 | Jamaica | JAM | Upper middle income | 3 | 0,42 | 0,4 | 3 | Latin America & Caribbean | 4 | Middle emissions |
| 2021 | Japan | JPN | High income | 4 | 0,37 | 0,69 | 3 | East Asia & Pacific | 6 | High emissions |
| 20 21 | Jordan | JOR | Upper middle income | 3 | 0,36 | 0,4 | 3 | Middle East & North Africa | 1 | Low emissi ons |
| 20 23 | Kazakhsta n | KAZ | Upper middle income | 3 | 0,32 | 0,51 | 3 | Europe & Central Asia | 3 | High emissi ons |
| 20 22 | Kiribati | KIR | Lower middle income | 2 | | 0,44 | 3 | East Asia & Pacific | 6 | |
| 20 21 | Korea Republic | KOR | High income | 4 | 0,37 | 0,72 | 3 | East Asia & Pacific | 6 | High emissi ons |
| 20 21 | Kuwait | KWT | High income | 4 | 0,35 | 0,46 | 3 | Middle East & North Africa | 1 | High emissi ons |
| 20 21 | Kyrgyz Republic | KGZ | Lower middle income | 2 | 0,33 | 0,39 | 3 | Europe & Central Asia | 3 | Low emissi ons |
| 20 21 | Lao PDR | LAO | Lower middle income | 2 | 0,46 | 0,33 | 1 | East Asia & Pacific | 6 | |
| 20 20 | Lebanon | LBN | Upper middle income | 3 | 0,4 | 0,28 | 2 | Middle East & North Africa | 1 | Middle emissi ons |
| 20 17 | Lesotho | LSO | Lower middle income | 2 | 0,47 | 0,3 | 1 | Sub-Saharan Africa | 5 | Low emissi ons |
| 20 21 | Liberia | LBR | Low income | 1 | 0,6 | 0,28 | 1 | Sub-Saharan Africa | 5 | Low emissi ons |
| 20 17 | Liechtenst ein | LIE | High income | 4 | | 0,63 | 3 | Europe & Central Asia | 3 | |
| 20 16 | Madagasc ar | MDG | Low income | 1 | 0,55 | 0,26 | 1 | Sub-Saharan Africa | 5 | Low emissi ons |

| 2021 | Malawi | MWI | Low income | 1 | 0,54 | 0,29 | 1 | Sub-Saharan Africa | 5 | Low emissions |
| 2021 | Malaysia | MYS | Upper middle income | 3 | 0,36 | 0,5 | 3 | East Asia & Pacific | 6 | High emissions |
| 2020 | Maldives | MDV | Upper middle income | 3 | 0,53 | 0,44 | 4 | South Asia | 7 | Middle emissions |
| 2021 | Mauritania | MRT | Lower middle income | 2 | 0,55 | 0,35 | 1 | Sub-Saharan Africa | 5 | Middle emissions |
| 2021 | Mauritius | MUS | High income | 4 | 0,42 | 0,56 | 3 | Sub-Saharan Africa | 5 | Middle emissions |
| 2022 | Mexico | MEX | Upper middle income | 3 | 0,38 | 0,36 | 2 | Latin America & Caribbean | 4 | Middle emissions |
| 2022 | Micronesia | FSM | Lower middle income | 2 | 0,61 | 0,35 | 1 | East Asia & Pacific | 6 | |
| 2020 | Moldova | MDA | Lower middle income | 2 | 0,4 | 0,44 | 3 | Europe & Central Asia | 3 | Low emissions |
| 2020 | Monaco | MCO | High income | 4 | | 0,75 | 3 | Europe & Central Asia | 3 | |
| 2020 | Mongolia | MNG | Lower middle income | 2 | 0,37 | 0,45 | 3 | East Asia & Pacific | 6 | High emissions |
| 2021 | Montenegro | MNE | Upper middle income | 3 | 0,36 | 0,46 | 3 | Europe & Central Asia | 3 | Middle emissions |
| 2021 | Morocco | MAR | Lower middle income | 2 | 0,37 | 0,42 | 3 | Middle East & North Africa | 1 | Low emissions |

| 2021 | Mozambique | MOZ | Low income | 1 | 0,49 | 0,26 | 1 | Sub-Saharan Africa | 5 | Low emissions |
|------|------------|-----|------------|---|------|------|---|--------------------|---|---------------|
| 2021 | Myanmar | MMR | Lower middle income | 2 | 0,5 | 0,25 | 1 | East Asia & Pacific | 6 | Middle emissions |
| 2021 | Namibia | NAM | Upper middle income | 3 | 0,46 | 0,38 | 1 | Sub-Saharan Africa | 5 | Middle emissions |
| 2021 | Nauru | NRU | High income | 4 | 0,58 | 0,49 | 4 | East Asia & Pacific | 6 | |
| 2020 | Nepal | NPL | Lower middle income | 2 | 0,49 | 0,36 | 1 | South Asia | 7 | Low emissions |
| 2021 | New Zealand | NZL | High income | 4 | 0,29 | 0,7 | 3 | East Asia & Pacific | 6 | High emissions |
| 2020 | Nicaragua | NIC | Lower middle income | 2 | 0,44 | 0,27 | 1 | Latin America & Caribbean | 4 | Middle emissions |
| 2021 | Niger | NER | Low income | 1 | 0,63 | 0,34 | 1 | Sub-Saharan Africa | 5 | Low emissions |
| 2021 | Nigeria | NGA | Lower middle income | 2 | 0,48 | 0,25 | 1 | Sub-Saharan Africa | 5 | Low emissions |
| 2016 | Niue | NIU | High income | 4 | | | 2 | East Asia & Pacific | 6 | |
| 2021 | North Macedonia | MKD | Upper middle income | 3 | 0,37 | 0,46 | 3 | Europe & Central Asia | 3 | High emissions |
| 2022 | Norway | NOR | High income | 4 | 0,26 | 0,76 | 3 | Europe & Central Asia | 3 | High emissions |
| 20 21 | Oman | OMN | High income | 4 | 0,41 | 0,5 | 3 | Middle East & North Africa | 1 | High emissi ons |
| 20 21 | Pakistan | PAK | Lower middle income | 2 | 0,52 | 0,31 | 1 | South Asia | 7 | Low emissi ons |
| 20 15 | Palau | PLW | High income | 4 | 0,48 | 0,42 | 4 | East Asia & Pacific | 6 | |
| 20 20 | Papua New Guinea | PNG | Lower middle income | 2 | 0,54 | 0,28 | 1 | East Asia & Pacific | 6 | High emissi ons |
| 20 22 | Paraguay | PRY | Upper middle income | 3 | 0,37 | 0,33 | 2 | Latin America & Caribbean | 4 | High emissi ons |
| 20 21 | Peru | PER | Upper middle income | 3 | 0,41 | 0,39 | 3 | Latin America & Caribbean | 4 | Middle emissi ons |
| 20 21 | Philippines | PHL | Lower middle income | 2 | 0,46 | 0,33 | 1 | East Asia & Pacific | 6 | Low emissi ons |
| 20 21 | Qatar | QAT | High income | 4 | 0,36 | 0,54 | 3 | Middle East & North Africa | 1 | High emissi ons |
| 20 20 | Russian Federation | RUS | Upper middle income | 3 | 0,32 | 0,54 | 3 | Europe & Central Asia | 3 | High emissi ons |
| 20 20 | Rwanda | RWA | Low income | 1 | 0,52 | 0,42 | 4 | Sub-Saharan Africa | 5 | Low emissi ons |
| 20 21 | Saint Lucia | LCA | Upper middle income | 3 | 0,39 | 0,45 | 3 | Latin America & Caribbean | 4 | Low emissi ons |
| 20 21 | Samoa | WSM | Upper middle income | 3 | 0,5 | 0,43 | 4 | East Asia & Pacific | 6 | Low emissi ons |

| | | | | | | | | | | | |
|---|---|---|---|---|---|---|---|---|---|---|---|
| 20 15 | San Marino | SMR | High income | 4 | | 0,63 | 3 | Europe & Central Asia | 3 | |
| 20 21 | São Tomé and Principe | STP | Lower middle income | 2 | 0,51 | 0,36 | 1 | Sub-Saharan Africa | 5 | Low emissi ons |
| 20 21 | Saudi Arabia | SAU | High income | 4 | 0,4 | 0,54 | 3 | Middle East & North Africa | 1 | High emissi ons |
| 20 20 | Senegal | SEN | Lower middle income | 2 | 0,51 | 0,34 | 1 | Sub-Saharan Africa | 5 | Low emissi ons |
| 20 22 | Serbia | SRB | Upper middle income | 3 | 0,41 | 0,44 | 3 | Europe & Central Asia | 3 | Middle emissi ons |
| 20 21 | Seychelles | SYC | High income | 4 | 0,45 | 0,47 | 4 | Sub-Saharan Africa | 5 | |
| 20 21 | Sierra Leone | SLE | Low income | 1 | 0,56 | 0,3 | 1 | Sub-Saharan Africa | 5 | Low emissi ons |
| 20 22 | Singapore | SGP | High income | 4 | 0,37 | 0,8 | 3 | East Asia & Pacific | 6 | High emissi ons |
| 20 21 | Solomon Islands | SLB | Lower middle income | 2 | 0,59 | 0,39 | 4 | East Asia & Pacific | 6 | High emissi ons |
| 20 21 | Somalia | SOM | Low income | 1 | 0,67 | 0,35 | 1 | Sub-Saharan Africa | 5 | Low emissi ons |
| 20 21 | South Africa | ZAF | Upper middle income | 3 | 0,38 | 0,35 | 2 | Sub-Saharan Africa | 5 | Middle emissi ons |
| 20 21 | Sri Lanka | LKA | Lower middle income | 2 | 0,46 | 0,39 | 4 | South Asia | 7 | Low emissi ons |

| 2021 | St. Kitts and Nevis | KAN | High income | 4 | 0,46 | 0,55 | 4 | Latin America & Caribbean | 4 | |
| 2015 | St. Vincent and the Grenadines | VCT | Upper middle income | 3 | | 0,47 | 3 | Latin America & Caribbean | 4 | Low emissions |
| 2021 | State of Palestine | PSE | Lower middle income | 2 | | | 2 | Middle East & North Africa | 1 | |
| 2021 | Sudan | SDN | Low income | 1 | 0,6 | 0,26 | 1 | Sub-Saharan Africa | 5 | Low emissions |
| 2020 | Suriname | SUR | Upper middle income | 3 | 0,39 | 0,33 | 2 | Latin America & Caribbean | 4 | High emissions |
| 2021 | Switzerland | CHE | High income | 4 | 0,24 | 0,69 | 3 | Europe & Central Asia | 3 | Middle emissions |
| 2018 | Syrian Arab Republic | SYR | Low income | 1 | 0,46 | 0,22 | 1 | Middle East & North Africa | 1 | |
| 2021 | Tajikistan | TJK | Low income | 1 | 0,37 | 0,32 | 2 | Europe & Central Asia | 3 | Low emissions |
| 2021 | Tanzania | TZA | Lower middle income | 2 | 0,5 | 0,3 | 1 | Sub-Saharan Africa | 5 | Low emissions |
| 2022 | Thailand | THA | Upper middle income | 3 | 0,43 | 0,48 | 4 | East Asia & Pacific | 6 | Middle emissions |
| 2022 | Timor-Leste | TLS | Lower middle income | 2 | 0,5 | 0,37 | 1 | East Asia & Pacific | 6 | High emissions |
| 2021 | Togo | TGO | Low income | 1 | 0,49 | 0,35 | 1 | Sub-Saharan Africa | 5 | Low emissions |
| 20 20 | Tonga | TON | Upper middle income | 3 | 0,6 | 0,42 | 4 | East Asia & Pacific | 6 | Low emissi ons |
| 20 18 | Trinidad and Tobago | TTO | High income | 4 | 0,36 | 0,33 | 2 | Latin America & Caribbean | 4 | High emissi ons |
| 20 21 | Tunisia | TUN | Lower middle income | 2 | 0,38 | 0,43 | 3 | Middle East & North Africa | 1 | Low emissi ons |
| 20 23 | Turkey | TUR | Upper middle income | 3 | 0,35 | 0,48 | 3 | Europe & Central Asia | 3 | Middle emissi ons |
| 20 22 | Turkmenis tan | TKM | Upper middle income | 3 | 0,34 | 0,23 | 2 | Europe & Central Asia | 3 | High emissi ons |
| 20 22 | Tuvalu | TUV | Upper middle income | 3 | | 0,61 | 3 | East Asia & Pacific | 6 | |
| 20 22 | Uganda | UGA | Low income | 1 | 0,58 | 0,28 | 1 | Sub-Saharan Africa | 5 | Low emissi ons |
| 20 21 | Ukraine | UKR | Lower middle income | 2 | 0,36 | 0,42 | 3 | Europe & Central Asia | 3 | Low emissi ons |
| 20 23 | United Arab Emirates | ARE | High income | 4 | 0,37 | 0,58 | 3 | Middle East & North Africa | 1 | High emissi ons |
| 20 22 | United Kingdom | GBR | High income | 4 | 0,28 | 0,68 | 3 | Europe & Central Asia | 3 | Middle emissi ons |
| 20 21 | United States | USA | High income | 4 | 0,3 | 0,65 | 3 | North America | | High emissi ons |
| 20 22 | Uruguay | URY | High income | 4 | 0,37 | 0,5 | 3 | Latin America & Caribbean | 4 | High emissi ons |
| 20 21 | Uzbekista n | UZB | Lower middle income | 2 | 0,36 | 0,4 | 3 | Europe & Central Asia | 3 | Middle emissi ons |
| 20 21 | Vanuatu | VUT | Lower middle income | 2 | 0,55 | 0,38 | 1 | East Asia & Pacific | 6 | Low emissi ons |
| 20 23 | Vatican City State | | High income | 4 | | | 2 | Europe & Central Asia | 3 | |
| 20 21 | Venezuela | VEN | Upper middle income | 3 | 0,38 | 0,18 | 2 | Latin America & Caribbean | 4 | Middle emissi ons |
| 20 22 | Vietnam | VNM | Lower middle income | 2 | 0,47 | 0,42 | 4 | East Asia & Pacific | 6 | Middle emissi ons |
| 20 21 | Zambia | ZMB | Lower middle income | 2 | 0,47 | 0,32 | 1 | Sub-Saharan Africa | 5 | Low emissi ons |
| 20 21 | Zimbabwe | ZWE | Lower middle income | 2 | 0,5 | 0,21 | 1 | Sub-Saharan Africa | 5 | Middle emissi ons |

**Prompt versions**

<u>First one (SDG assignation):</u>
"Assign the following text to all relevant SDGs (strictly from the following list: 1) No poverty 2) Zero Hunger 3) Good health and well-being 4) Quality education 5) Gender equality 6) Clean Water and Sanitation 7) Affordable and clean energy 8) Decent work and economic growth 9) Industry, innovation and infrastructure 10) Reduced inequalities 11) Sustainable cities and communities 12) Responsible consumption and production 13) Climate action 14) Life below water 15) Life on land 16) Peace, justice and strong institutions 17) Partnerships for the goals). If a paragraph tackles non relevant issues with respect to any SDG, assign 0."

<u>Second prompt_ option 1</u>
"Consider climate adaptation as the adjustment in natural or human systems in response to actual or expected climatic stimuli or their effects, which moderates harm or exploits beneficial opportunities. Also consider, climate mitigation as an anthropogenic intervention to reduce the sources or enhance the sinks of greenhouse gases. For each paragraph in the file assign 0 if the paragraph explains or present an action or a set of actions which pose risks to at least one between climate adaptation and mitigation; assign 1 if the paragraph is neutral with respect to climate adaptation and mitigation and does not express or discuss any judgement; assign 2 if the paragraph explains or present an action or a set of actions which pose opportunities to at least one between climate adaptation and mitigation"

<u>Second prompt_ option 2</u>
"Consider climate adaptation as the adjustment in natural or human systems in response to actual or expected climatic stimuli or their effects, which moderates harm or exploits beneficial opportunities. Also consider, climate mitigation as an anthropogenic intervention to reduce the sources or enhance the sinks of greenhouse gases. Assign to each paragraph one and one only number between 0, 1 and 2. Assign 0 if if the paragraph explains or present an action or a set of actions which pose risks to at least one between climate adaptation and mitigation; assign 1 if the paragraph is neutral with respect to climate adaptation and mitigation and does not express or discuss any judgement; assign 2 if the paragraph explains or present an action or a set of actions which pose opportunities to at least one between climate adaptation and mitigation"

<u>Second prompt_option 3</u>
"Consider climate adaptation as the adjustment in natural or human systems in response to actual or expected climatic stimuli or their effects, which moderates harm or exploits beneficial opportunities. Also consider, climate mitigation as an anthropogenic intervention to reduce the sources or enhance the sinks of greenhouse gases. For every paragraph [i] in '{paper_text}', assign a unique value between 0, 1 and 2 respecting the following rule: Assign 0 if if the paragraph explains or present an action or a set of actions which pose risks to at least one between climate adaptation and mitigation; assign 1 if the paragraph is neutral with respect to climate adaptation and mitigation and does not express any judgement; assign 2 if the paragraph explains or present an action or a set of actions which create opportunities to at least one between climate adaptation and mitigation. Restrict your outcome to one digit only, without motivating your choice with text".

<u>Second prompt_option 4</u>
"Consider climate adaptation as the adjustment in natural or human systems in response to actual or expected climatic stimuli or their effects, which moderates harm or exploits beneficial opportunities. Also consider, climate mitigation as an anthropogenic intervention to reduce the sources or enhance the sinks of greenhouse gases. For every paragraph [i] in '{paper_text}', assign a unique value between 0, 1 and 2 respecting the following rule: Assign 0 if if the paragraph explains or present an action or a set of actions which pose risks to at least one between climate adaptation and mitigation; assign 1 if the paragraph is neutral with respect to climate adaptation and mitigation and does not express any judgement; assign 2 if the paragraph explains or present an action or a set of actions which create opportunities to at least one between climate adaptation and mitigation. Restrict your outcome to one digit only, without motivating your choice with text".

<u>Second prompt_ option 5</u>

Use the following rules to interpret a paragraph: Consider climate adaptation as the adjustment in natural or human systems in response to actual or expected climatic stimuli or their effects, which moderates harm or exploits beneficial opportunities. Also consider, climate mitigation as an anthropogenic intervention to reduce the sources or enhance the sinks of greenhouse gases. Assign to each paragraph one and one only number between 0, 1 and 2. Assign 0 if if the paragraph explains or present an action or a set of actions which pose risks to at least one between climate adaptation and mitigation; assign 1 if the paragraph is neutral with respect to climate adaptation and mitigation and does not express or discuss any judgement; assign 2 if the paragraph explains or present an action or a set of actions which pose opportunities to at least one between climate adaptation and mitigation.